\documentclass[preprint]{aastex}
\usepackage{natbib}
\usepackage{graphicx}
\usepackage{subfigure}
\usepackage{amsmath,amssymb}
\begin{document}

\title{Long Duration Flare Emission: Impulsive Heating or Gradual Heating?}
\author{Jiong Qiu \& Dana W. Longcope}
\affil{Department of Physics, Montana State University, Bozeman MT 59717-3840, USA}

%
%
%
\begin{abstract}

Flare emissions in X-ray and EUV wavelengths have previously been modeled as the plasma 
response to impulsive heating from magnetic reconnection. 
Some flares exhibit gradually evolving X-ray and EUV light curves, which are believed
to result from superposition of an extended sequence of impulsive heating events occurring in 
different adjacent loops or even unresolved threads within each loop. In this paper,
we apply this approach to a long duration two-ribbon flare SOL2011-09-13T22 observed by the 
{\em Atmosphere Imaging Assembly} (AIA). We find that, to reconscile with observed signatures 
of flare emission in multiple EUV wavelengths, each thread should be heated in two phases, an intense 
impulsive heating followed by a gradual, low-rate heating tail that is attenuated over 20-30 min.
Each AIA resolved single loop may be composed of several such threads.
The two-phase heating scenario is supported by modeling with both a 0d and 
a 1d hydrodynamic code. We discuss viable physical mechanisms for the two-phase heating in a 
post-reconnection thread.

\end{abstract}

\section{Introduction}
A solar flare is characterized by increased radiation across a large domain of the electromagnetic spectrum, which
has been observed for a few decades with generations of instrumentation. 
Based on these observations, the general picture is agreed upon that flare plasmas, whether 
in the corona or in the lower-atmosphere, are heated on relatively short timescales.
As is commonly accepted, energy release in flares is governed by magnetic reconnection 
in the corona on Alfv\'enic timescales of order a few seconds \citep{Priest2002}. 
Although the debate exists regarding where exactly in the Sun's atmosphere 
particles or plasmas are energized primarily, what is the form of heating, and how energy is transferred between different layers of the atmosphere, 
the impulsive rise of flare emission in many wavelengths, including hard X-ray, microwave, optical, and UV
bands, is considered to reflect short timescales of flare energy release and also heating \citep{Fletcher2011}. 
The flare plasmas in the corona then cool down by conduction and radiation \citep{Culhane1970, Antiochos1978,
Cargill1995}. Hydrodynamic flows also play a crucial role in heating or cooling the corona by means of enthalpy flows in different phases of the flare evolution, typically chromospheric evaporation in the early 
heating phase \citep{Fisher85_6, Fisher87, Longcope2014} that both fills and heats the corona, and coronal condensation in 
the cooling phase \citep{Bradshaw2010a, Bradshaw2010b}, and these processes take place on acoustic timescales
or shorter.

Observations have shown that soft X-ray and EUV emissions in many
flares appear to evolve and decay more slowly than cooling timescales, if 
only a one-time impulsive heating is introduced in the rise phase of the flare. 
It is therefore considered that the gradual or decay phase of a flare, after the impulsive rise of its emission, 
cannot be solely governed by cooling, but additional heating has to be invoked. With observations
obtained by missions from as early as {\em Skylab} to the more recent {\it Reuven Ramaty 
High Energy Solar Spectroscopic Imager} \citep[{\it RHESSI};][]{Lin2002}, the {\it Atmospheric Imaging Assembly} 
\citep[{\it AIA};][]{Lemen2012}, and {\it Extreme Ultraviolet Variability Experiment} \citep[EVE;][]{Woods2012},
it has been noted that, for large flares, a significant amount of heating energy could be provided
during the decay phase, sometimes more than during the rise phase \citep{Withbroe1978, Dere1979, 
Jiang2006, Sun2013, Ryan2013}. \citet{Dere1979}, observing the movement of X-ray emission source, 
put forward that the continuous heating into the decay phase is in the form of ``sequential heating of 
new loops in the flare region". The same insight was given by \citet{Antiochos1980} studying cooling 
of flare plasmas theoretically. This scenario is firmly supported by modern observations with high spatial resolutions, such as the {\it Transition Region and Coronal Explorer} \citep[{\it TRACE};][]{Handy1999},
revealing that a flare comprises a multitude of plasma loops formed and heated independently 
at different times during its evolution \citep{Aschwanden2001, Aschwanden2001a}. 

With this idea, \citet{Hori1997} applied a 1d hydrodynamic loop model to a 
stack of loops heated successively with a prescribed heating rate 
to simulate flare soft X-ray emission. Following their effort, a few studies modeling successive heating of flare loops have been attempted in order to reproduce elongated flare emission
\citep{Reeves2002, Warren2006, Longcope2010, Hock2012, Qiu2012, Liu2013}. 
These multi-loop models have approached the problem along different avenues, using different
methods and forms of initial inputs, yet all with the common final goal to match the 
synthetic and observed soft X-ray and/or EUV light curves. \citet{Reeves2002} modeled the 
post-flare arcade of the Bastille-day flare on 2000 July 14 by launching 500 loops, each 
starting 4~s after the previous loop. Instead of modeling the heating process, a scaling law 
was used to evolve plasma temperature and density from given initial values, which were
adjusted to best fit the soft X-ray and EUV 195\AA\ light curves observed by Soft X-ray Telescope \citep[SXT; ][]{Tsuneta1991} 
and TRACE, respectively. \citet{Warren2006} ran a 1d hydrodynamic simulation to model the Masuda flare on 1992 January 13 \citep{Masuda1994}.
In this model, 50 loops were introduced with a 40~s interval and 200~s heating duration, and the amount
of heating energy in each loop was different and determined from the observed GOES soft X-ray flux. The model
synthetic light curves compared favorably with soft X-ray observations by SXT and the {\it Bragg Crystal Spectrometer} \citep[BCS; ][]{Culhane1991}.
\citet{Longcope2010} developed a model of flare heating by reconnection and compression, and used
observationally inferred reconnection flux in successively formed flare loops
to derive post-reconnection plasma properties and synthesize the super-hot loop-top source of an X-class flare
on 2004 February 26 observed by RHESSI. Very recently, \citet{Liu2013} used spatially resolved UV 
1600\AA\ lightcurves of a flare on 2005 May 13 observed by TRACE to identify foot-points of successively heated
flare loops and infer heating profiles in over five thousand loops assumed to be anchored at these UV brightened
pixels. A 0d loop model \citep{Klimchuk2008} was then used to calculate mean plasma properties in these
loops and synthesize the flare soft X-ray spectrum and light curves observed by RHESSI and GOES. 
These multi-loop studies have exclusively synthesized flare X-ray emission at relatively high 
temperatures. In the era of the {\em Solar Dynamic Observatory}(SDO), \citet{Hock2012} and \citet{Qiu2012} 
were able to synthesize EUV light curves observed by AIA,
characterizing plasma temperatures from 10~MK down to 1~MK, and therefore also addressing cooling of 
successively heated multiple loops to the very late stage of the flare evolution.

Essential to all these models, heating of a flare loop is introduced as a short pulse with
a timescale of no more than a few minutes, echoing the prevailing belief that flare heating is primarily
impulsive. The long duration of the total flare emission has been believed to result from superposition of
multiple loops successively heated throughout the flare evolution. Whereas earlier studies used prescribed heating rates to model and match flare light curves, the most recent endeavors by \citet{Longcope2010, Qiu2012, Liu2013} 
have attempted to identify individual {\em impulsive} heating events from observations of flare foot-points, and confirmed that 
reconnection and heating proceed into the decay phase of the flare. However, these studies still 
cannot produce enough emission, raising the question whether the method has missed weak heating events or unresolved 
sub-structures, or the loop modeling is faulty, or the heating profile in these loops is 
different from what we have thought.

This paper tests each of the three scenarios above to shed light on
the invisible yet critical heating process of flare loops. We find that,
hydrodynamic models using a modified heating profile with an 
impulsive pulse followed by a gradual low-rate tail to 
heat each of multiple {\em threads} in a single {\em loop} may 
reconscile with observed flare emission signatures.\footnote{For clarity and consistency, in this paper, we
use the word {\em loops} to indicate loop like structures observed
in EUV images or inferred from foot-point 
emission in UV images, whereas the word {\em threads} 
refers to presumed substructures within a loop that are not resolved in images. 
A {\em heating event} refers to heating in a loop or a thread, depending on the context.}
In the following text, we demonstrate these experiments applied to a 
long duration C-class two-ribbon flare observed by AIA, starting with a 
method developed by \citet{Qiu2012} and \citet{Liu2013} to 
identify heating events during the flare. The method uses spatially 
resolved UV emission, which is assumed to be from the feet 
of flare loops, to infer heating rates in these loops, and 
is hereafter called the UV Footpoint Calorimeter (UFC) method. 
The flare exhibits an arcade of loops formed sequentially along 
the magnetic polarity inversion line and evolving slowly from 10 - 1 MK, as described in Section 2. Section 3 
shows that analysis and modeling with the standard UFC method can reproduce the global evolution pattern of 
the flare EUV emission in the arcade but with large discrepancies in evolution of single loop pixels. In Section 4, 
modified UFC method and 0d modeling are applied treating a single-loop as a 
cluster of unresolved sub-structures or threads, and different modulations of heating rates are attempted in
these threads in order to resolve discrepancies with observations. In Section 5, simulation with a 1d hydrodynamic code
is performed to compare with the results by the 0d model. Conclusions and discussions 
are presented in the last section.

\section{Long Duration Flare Emission}
\subsection{Overview of the Flare}
A small C-class two-ribbon flare occurred at 22 UT on 2011 September 13 in the active region
AR11289. The active region is characterized by a bipolar magnetic configuration consisting of plage of positive
polarity and a sunspot of negative polarity. The flare is associated with a coronal mass ejection,
and is a long duration event with prolonged thermal emission lasting more than four hours.
The flare took place near the disk center and was observed by AIA onboard SDO.

\begin{figure}
\epsscale{1.0}
\plotone{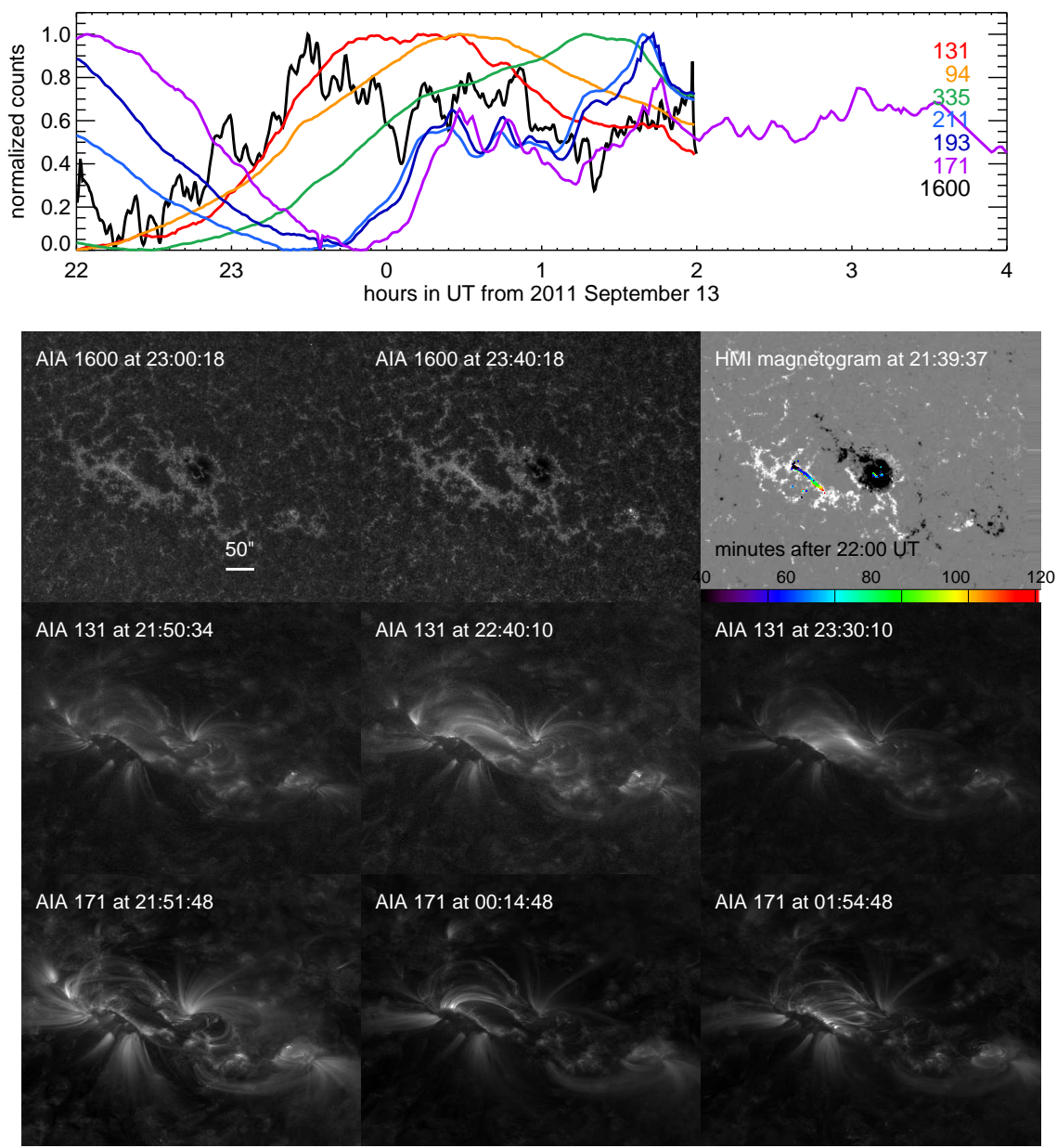}
\caption{Top: Light curves of total UV and EUV counts in the flaring active region observed by AIA in UV 1600\AA, EUV 131, 94, 335, 211, 193
and 171\AA, respectively. Pre-flare emission is subtracted off the light curve, which is then normalized to its maximum.
Bottom: images of the flare observed by AIA at UV 1600 \AA\ showing the flare ribbon evolution (top), and in EUV 131 \AA\ (middle) 
and 171 \AA\ (bottom) showing evolution of the flare arcade. The top-right panel shows a longitudinal magnetogram obtained by HMI 
superimposed with positions of the brightest ribbon pixels that are brightened sequentially at different times indicated 
by the color code.} \label{overview}
\end{figure}

Figure~\ref{overview} shows time evolution of the total counts within an area including all flare loops 
from six AIA EUV bands at 131, 94, 335, 211, 193, and 171 \AA, illustrating
thermal emission at characteristic plasma temperatures from 10 to 1~MK. 
Also shown is the total count light curve in AIA UV 1600 band, which
captures enhanced C{\sc iv} emission at 100,000~K during the flare, as well as nearby continuum. 
The flare first exhibits enhanced UV 1600 emission at locations of the feet of flare loops (see images in the lower panels). The UV emission is then followed by coronal temperature 
emissions, 10 MK in 131\AA, 6 MK in 94\AA, 3~MK in 335\AA, and finally 1-2 MK in 211/193/171\AA. 
Such a sequence corroborates the standard flare model, depicting that energy deposition into the lower atmosphere
drives plasmas into newly formed coronal loops (chromopsheric evaporation) and heated to at least 10 MK, and these loops then cool down to a few million degrees giving rise to EUV emissions at subsequently lower temperatures.

The figure also shows images of the flare in a few bands, confirming the above-mentioned scenario of flare
evolution. In UV 1600 band, two patches are brightened simultaneously in positive and 
negative magnetic fields, respectively, making conjugate foot-points of an 
arcade of flare loops that are brightened afterwards. Notably in this event, the UV 
brightening most evident in one ribbon spreads slowly from northeast to 
southwest along the ribbon, nearly parallel to the magnetic polarity inversion 
line (PIL), at roughly $v\simeq 15$ km s$^{-1}$, as outlined in Figure~\ref{overview}.
This ribbon elongation is followed by the same slow, orderly progression of loop brightenings,
first in 131 images characteristic of higher temperatures, and then in 
subsequently lower temperatures in 94, 335, and then 211/193/171 bands.
It is hence evident from these observations that reconnection and formation of 
flare loops take place sequentially in a generally organized manner. 

RHESSI observed the flare from its rise to peak in 3 - 12 keV.
Light curves of the flare at these photon energies exhibit a rather gradual
evolution. 
From these observations, it is likely that thick-target non-thermal 
emission is insignificant in this flare. 
Spatially unresolved soft X-ray and EUV emissions 
are also obtained by GOES and EVE, respectively. However, another flare close to the 
eastern limb 800\arcsec\ away took place during the peak of the target flare. Since 
it is difficult to separate different emission components by the two flares, observations 
by these instruments (RHESSI, GOES, and EVE) are not analyzed in this study.

\subsection{Evolution of Sequentially Formed Flare Loops}
The overall organization shown in this flare provides a good opportunity to
study the formation of flare loops, and heating and cooling of these loops. 
We proceed by assuming that each UV-brightened pixel is the footpoint of a heated flare loop which subsequently 
cools into the various EUV passbands. The foot-point UV emission, which is enhanced within 
seconds of energy deposition, allows us to track the sequence of reconnection energy release.  The evolution 
of post-flare loops observed in different wavelengths provides data on the cooling process.

\begin{figure}
\epsscale{1.0}
\plotone{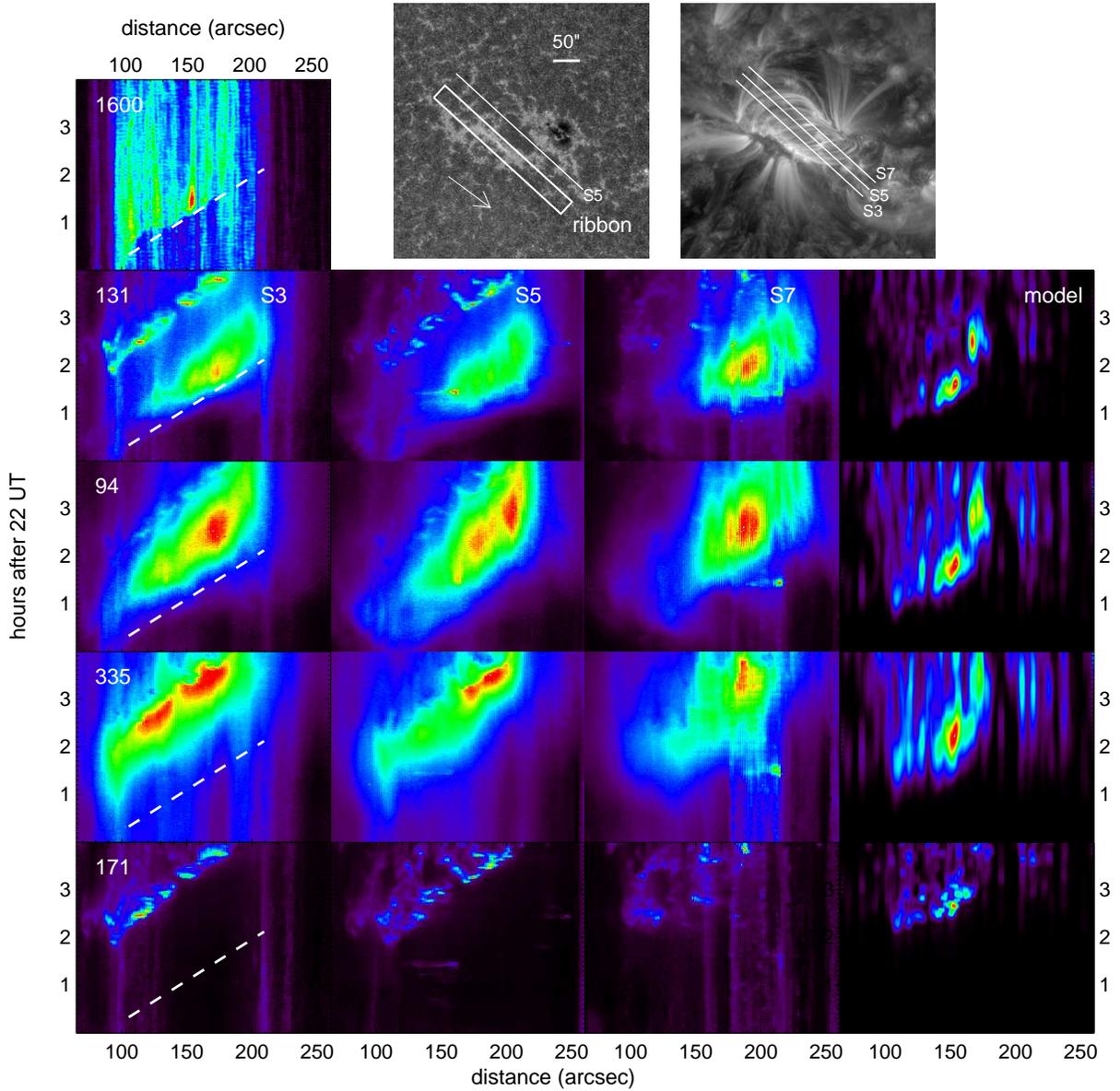}
\caption{Time-distance stack plots of flare emissions in UV 1600\AA\ along the ribbon (top left panel), 
and in EUV 131 (second row), 94 (third row), 335 (fourth row), and 171 \AA\ (bottom row) bands along three 
slits S3 (left column), S2 (second column), and S3 (third column) across the arcade. The ribbon 
emissions are derived as the mean intensity across the width of the rectangle along its length, from 
northeast to southwest, as denoted in the top middle panel, and the loop emissions are read off 
along the slits crossing the arcade as shown in the top right panel. The dashed guide line in the 
left column marks the front of the UV ribbon brightening. The rightmost column shows the synthetic stack plots 
at 131, 94, 335, and 171~\AA\ bands, respectively, computed with the EBTEL model (see Section 3).}  \label{intensitygram}
\end{figure}

To capture the sequence of these loops, we define nine artificial slits each running parallel 
to the PIL. Three of these slits are shown in the top right panel of Figure~\ref{intensitygram}.  
Adjacent slits are separated by $10\arcsec$ NS, and are labeled S1 -- S9 from the positive legs 
to the loop tops and then the negative legs.  EUV intensity, in units of counts per second per pixel, is read off the pixels along a slit, and assembled into a time-distance 
stack plot, shown in Figure~\ref{intensitygram}. This is done for each of the 6 Fe-sensitive wavelengths of AIA, 
for each of the nine artificial slits. The loop footpoints are captured in UV 1600, for which we use a different, 
thicker slit, parallel to the other nine. This slit, 50 AIA pixels ($31\arcsec$) wide, is shown as a rectangular box 
in the middle top panel of Figure \ref{intensitygram}. The UV counts along this slit are averaged over its width.  
To construct the stack plots in the transformed coordinates as along and perpendicular to the slits, we interpolate 
observed data counts in a refined grid of 0.325\arcsec\ per pixel. 
The time cadence of the data used in the analyses is 24~s for UV 1600\AA\ observation and 1~min for EUV observations.

Figure~\ref{intensitygram} shows the time-distance stack plot of the UV ribbon (top left panel), as well as
stack plots of EUV emissions along three of the nine slits for four bandpasses at 131, 94, 335, 
and 171\AA. Because of the similarity in morphology and evolution of EUV emissions observed in 211, 193, and 171 bands,
we only present analyses of the 171 data. In these stack plots, time is measured from 22UT of 2011 
September 13, and distance is measured in arcseconds along the slit from its northeast end.
The dashed guideline, indicating a speed of $v \approx 15$ km s$^{-1}$, in the top left panel tracks the front of UV brightening along the ribbon, 
which is then superimposed in the stack plots of EUV loop emission in the panels below.  Note that post-flare 
loops have varying amounts of shear: loops formed later (further south) appear to be more sheared than earlier-formed loops. 
As a result, the arcade extends further south than does the ribbon.  This means that a given spatial coordinate in 
different slits may correspond to different loops.

The time-distance stack plots in Fig.\  \ref{intensitygram} corroborate the general pattern of the ribbon and arcade evolution, 
consistently confirming the sequence of energy release along the PIL and subsequent heating and cooling
of flare loops.  Each stack plot shows one or two bands of emission with a slope similar to that of the ribbon.
The vertical separation between an EUV front and the UV front (i.e.\ the dashed line) roughly measures the time lag between 
energy release, immediately after which foot-point UV emission is produced, and loop emission in that EUV passband.  
This time lag is a measure of the time taken for the plasma to cool to the characteristic temperature of that particular EUV 
bandpass. It appears that flare loops are quickly heated to emit in 131\AA\ band
at 10~MK. It then cools down to emit at 6~MK (94\AA\ ) in about 
20 minutes, at 3~MK (335~\AA\ ) within an hour, and then at 1~MK (171~\AA\ ) in about 1.5 hours. EUV emission along different slits exhibits varying distribution along the 
loops, yet reserves the same evolution sequence along the PIL.

\subsection{Timescales of EUV Emissions in Flare Loops}

\begin{figure}
\begin{minipage}[b]{1.0\textwidth}
\includegraphics[width=1.\textwidth]{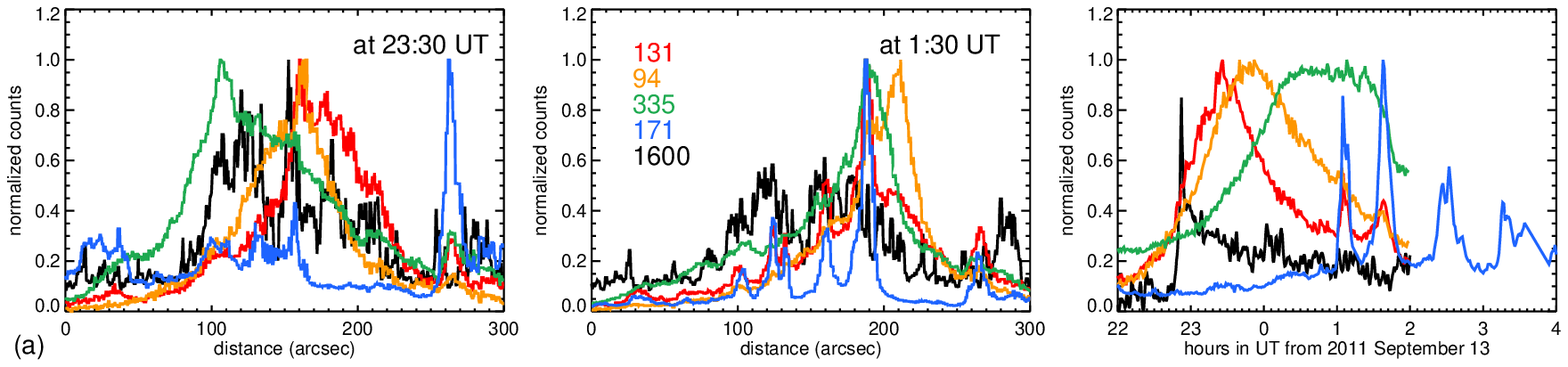}
\end{minipage}
\begin{minipage}[b]{1.0\textwidth}
\includegraphics[width=1.\textwidth,clip=]{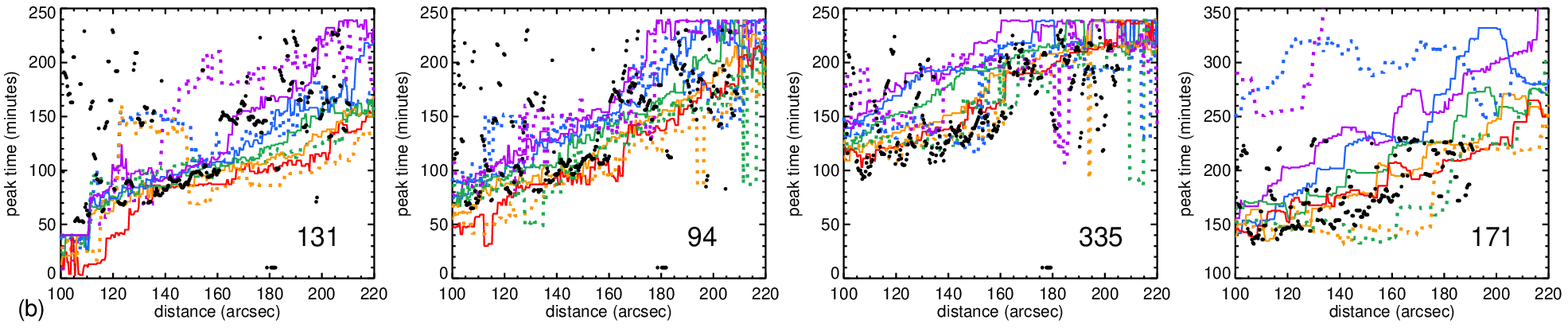}
\end{minipage}
\begin{minipage}[b]{1.0\textwidth}
\includegraphics[width=1.\textwidth,clip=]{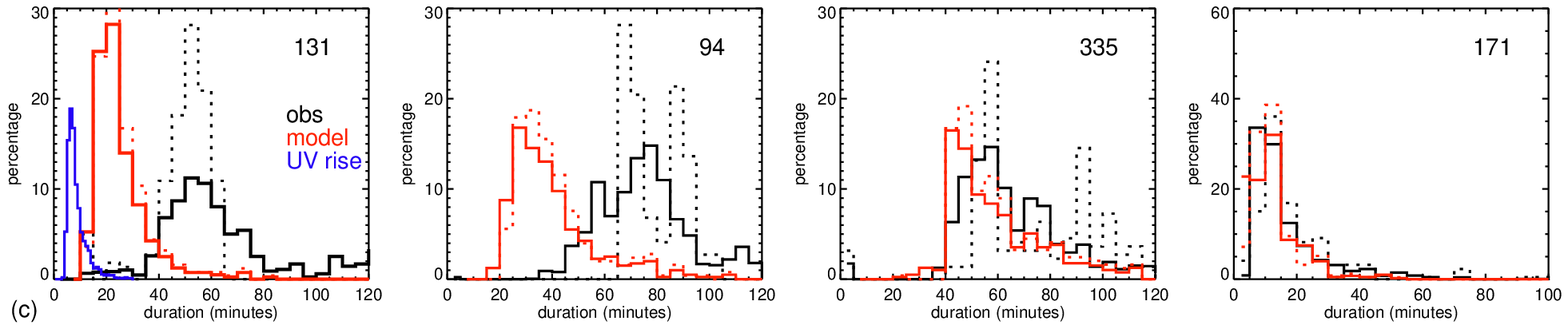}
\end{minipage}
\begin{minipage}[b]{1.0\textwidth}
\includegraphics[width=1.\textwidth,clip=]{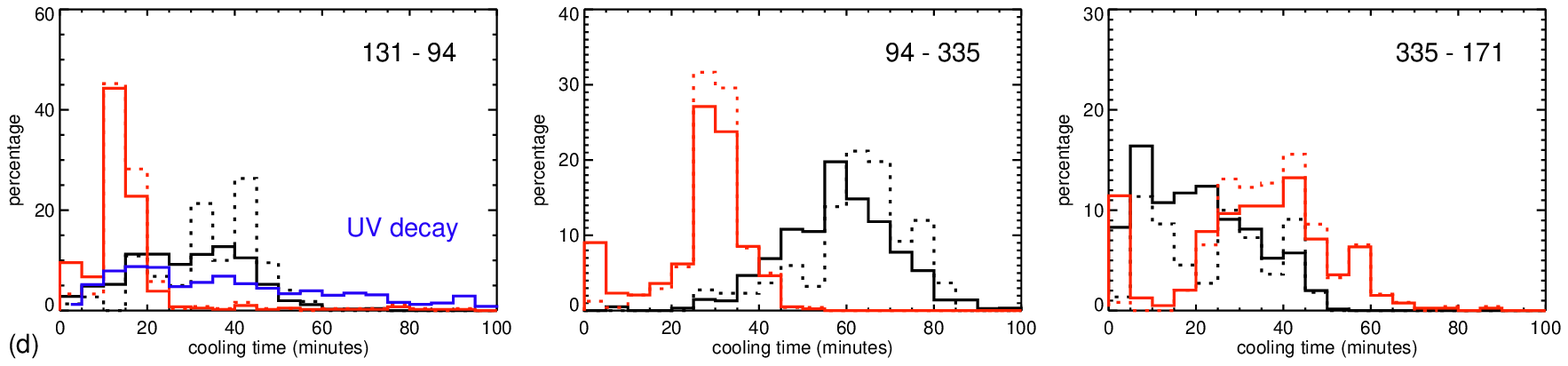}
\end{minipage}
\caption{Top (a): UV emission along the ribbon and EUV emissions along slit S5 at two times, the peak of the 
flare at 23:30 UT (left) and during the decay phase at 1:30 UT (middle), and time profiles of the UV emission 
at one location along the ribbon and EUV emissions at one location along slit S5 (right). 
Upper-middle (b): peak times of EUV emissions along 9 slits in 131, 94, 335, and 171 bands, respectively. 
In each plot, the red solid line shows the observed peak times along the central slit S5, and the
orange, green, blue, and violet solid (dotted) lines show the peak times along the slits to the left (right) of S5 toward
the feet of the arcade, respectively. The black dots show the peak times of the synthetic EUV 
emissions from the EBTEL modeling (see Section 3). In each plot, the peak time in minutes is counted from 22~UT.
Lower-middle (c): histograms of the duration of the observed (black) and synthetic (red) EUV emission in 4 AIA bands.
The histogram of the rise time of the UV 1600 lightcurves of the flaring pixels is also given in the left panel (blue).
Bottom (d): histograms of the cooling time of observed (black) and synthetic (red) EUV emission between different AIA bands.
Histogram of the $e$-slope decay time of the UV 1600 lightcurves of the flaring pixels is given in the left panel
(blue).}\label{stat}
\end{figure}

The stack plots can be further analyzed to find temporal and spatial scales of individual flare loops.
The UV 1600 stack plots show spatial structure of scales comparable with the instrument resolution 
\citep[~1-2\arcsec, ][]{Boerner2012}. UV emission is rapidly enhanced 
within a few minutes, presumably reflecting the short durations of the energy release episode in a given loop. 
In marked contrast to this, the EUV bands show variations over larger scales in both space and time.   
This contrast is even more obvious in horizontal and vertical slices of the stack-plots such as those shown in the 
top panels of Figure~\ref{stat} (Figure~\ref{stat}a).  The first two panels are horizontal slices sampling the entirety of slit S5 
during the peak of the flare (left), and during the decay of the flare (middle).  The 1600\AA\ and 171\AA\ bands 
exhibit distinct narrow spatial structures indicative of individual loops. In 171 \AA, the apparent loop 
width (FWHM) is typically of order a few arcseconds, the sharpest structure being 3 arcseconds. The 
other bandpasses, formed at higher temperatures, are more diffuse with scales of several tens of arcseconds. %
Later on in the decay phase (top middle panel of the figure), loops 
stand out in the 131 channel as much sharper structures with much 
of the wide envelop diminished.  These loops
are the same (cool) loops observed in the 171 band; they become visible in the 131 band because of the pronounced
response at 0.4 MK (Fe {\sc viii}) in this bandpass \citep{Odwyer2010}.
It is not clear what produces the wide envelop in high-temperature EUV emission in the early phase. We will further discuss 
this phenomenon in the next section.

Vertical slices of the stack plot, such as the top right panel of Figure~\ref{stat}, show the
time profiles from a single loop pixel. These can be used to estimate the duration of emission 
at a given bandpass as well as the cooling time between different bands. This particular panel shows
time profiles of the UV emission at one location on the ribbon as well as EUV emissions at one 
location along S5. Here the UV emission in bright kernels typically exhibits a rapid rise in a few minutes,
followed by a gradual decay of tens of minutes. EUV emissions in 131, 94, 335, and 171 bands rise, peak, and decay subsequently. 

The EUV time profiles of each pixel along each of the nine slits are analyzed to find timescales and delays. The time of peak emission at each bandpass is 
found for a given spatial location along a given slit. 
The second row of Figure\ \ref{stat} (Figure~\ref{stat}b) shows four panels corresponding to the peak times of the four passbands,
131, 94, 335 and 171\AA, respectively.  In each panel the peak times are plotted against slit position in a different color or line
style for each of the 9 slits. The red solid line shows the peak times along the central slit S5, and orange, 
green, blue, and violet solid (dotted) lines show the peak times along the
slits to the left (right) of S5 toward 
the feet of the arcade, respectively. 
The upward trends of these lines are further corroboration of the southward motion of 
the reconnection and arcade emission. This trend is less clear for the two slits plotted in red and orange (S8 and S9) 
in the 171\AA\ panel.  We henceforth focus attention on the other 7 slits in which complete evolution is observed in every wavelength.\footnote{The cadence of the EUV data we have processed and analyzed is one minute. To suppress spurious fluctuations, we also 
smooth the EUV time profile of each pixel with a 4-min box car; therefore, the timing accuracy in the following analyses 
is limited by these procedures to be no better than 4 minutes. This should not impact the analysis of observed evolution timescales, 
since these timescales are usually much longer than 4 minutes, as will be shown in the following.}

For further analyses, we select pixels at the distance between 
100\arcsec\ and 180\arcsec\ along slits S1--S7, for times between 22~UT on 2011 September 13 
and 2~UT on 2011 September 14. A total number of 1750 pixels of 0.325\arcsec\ by 0.325\arcsec\ in size are analyzed; seen in Figure~\ref{intensitygram}, 
these pixels make the central part of the arcade with relatively strong emissions.

For EUV emission at each pixel in each band, the duration is defined as the difference between the times of 
60\% of the peak emission in the rise and decay phase. Histograms of the duration for all the analyzed 
loop pixels in different bands are presented in the third row of Figure~\ref{stat} (Figure~\ref{stat}c)
and listed in Table 1 as well: on average, emission in 131, 94, and 335 bands each lasts for nearly an hour, and 
for only 10 minutes in 171 band. The median duration in these four bands is 66 $\pm$ 53, 74 $\pm$ 18, 60 $\pm$ 22,
and 12 $\pm$ 44 min, respectively, the uncertainties being the standard deviation. We note that the duration in 
171\AA\ band could be shorter than measured here, because of the smoothing window of 4-min we have applied to the 
light curves. We finish by noting that many of the pixels exhibit multiple peaks in the 171 band, from which our routine selects
the brightest peak for analysis; the other peaks typically have comparable durations.

We also measure the cooling time, defined as the time lag between peak emissions in two adjacent passbands, 
namely, the 131-94, 94-335, and 335-171 pairs. Histograms of the cooling time 
are shown along the bottom row of panels in Figure~\ref{stat} (Figure~\ref{stat}d). The median cooling time estimated this way 
is $27\pm 26, 58\pm 14$, and $18 \pm 19$~min between 131-94, 94-335, and 335-171 pairs, respectively. We also note 
that, in general, there is no significant difference (within 1$\sigma$) between histograms constructed along different 
slits, suggesting that loops evolve rather coherently. 
Note that the figures only present positive values of this time lag, or the ``cooling" time. In a small number of pixels
(5 - 20\% of the total, depending on the bandpass pairs), the time lag is negative between certain pairs. We consider the negative time lag
as due to uncertainties in identifying the peak emission in a very broad time profile or a time 
profile with multiple peaks. These negative time lags are not indications of the ``heating" time, because no single 
pixel exhibits negative lags in all three pairs.


The evolution through the highest temperatures is notably slow in this flare, as 
reflected in the long durations in 131, 94 and 335\AA\ (66, 74 and 60 mins, respectively) and the long 
delay between successive pairs (27 and 58 mins). Whereas the broad temperature responses of these AIA EUV bands may contribute to this,  
it is also probable that multiple loops cross the line-of-sight of any single pixel, and their convolved evolution will produce 
longer apparent duration. 
Overlapping loops are, however, rooted at distinct footpoints in the lower-atmosphere that brighten in UV emission, which can 
be distinguished or resolved to an extent limited by instrument's spatial resolution. 

In the following sections, we take the advantage of spatially resolved UV observations to identify loop
heating events, from which we compute the total emission by these loops to compare with the observed 
loop emissions and their properties. This method (UFC) assumes that each of 
the ribbon pixels is the foot-point of a single flare loop subject to a reconnection-related heating event, whose time 
profile and amplitude can be determined from the impulsive brightening at the footpoint observed in UV 1600. 
Furthermore, it is likely that such a ``single" {\em loop} consists of unresolved sub-structure, called 
{\em threads} \citep{Aschwanden2001,Warren2006}, which are heated at different times and with different amplitudes. 
In this case, we consider that the total energy deposited into all threads composing a single loop is constrained 
by the UV 1600 light curve at the foot of the loop, and conduct a few experiments modulating the frequency 
and amplitude of thread-heating events. Through these experiments, we find that the best match to observed coronal 
emission properties is provided by using a few threads, but each having a heating profile with a long, low-intensity tail. 
These experiments are described below.

\section{Zero-dimensional Modeling of 12,500 Loops}

We attempt to understand evolution of flare loops using the UFC method. The basics of this 
method were presented by \citet{Qiu2012, Liu2013}. 
The standard UFC method 
assumes that a single flare {\em loop} is rooted at a ribbon pixel brightened in UV 1600 bandpass, and this ribbon pixel 
is hereafter called the foot-point of the flare loop. We identify these loop footpoints from the $0.325\arcsec$ 
pixels in the 325\arcsec\ (along the slit) by 31\arcsec\ (across the slit) rectangular box enclosing the positive flare ribbon 
(see Figure\ \ref{intensitygram}. We dismiss UV data from the negative ribbon because of the lower signal-to-noise ratio there).
A single pixel is identified as a loop foot-point if its emission increases to at least twice the pre-flare level for more than 5 minutes.  
A total of 12,500 pixels (about 13\% of all pixels in the box), and therefore 12,500 single flare {\em loops}, are thus 
identified at different times during the 4 hours of observation.  

The plasma evolution of a single loop is computed using the zero-dimensional (0d) EBTEL model of 
\citet{Klimchuk2008}. All loops are given a half-length $L =  57.3$~Mm, which is $\pi/2$ times 
the mean distance between the positive and negative UV ribbons. The initial rise of the pixel's UV 1600
curve is fit to a half-Gaussian to determine its peak time, $t_0$, peak intensity $I_0$, and rise time $\tau_0$.  
The EBTEL model is run with an asymmetric heating profile based on these values, 
\begin{equation}
H(t) = \left\{ \begin{array}{lcl} \lambda I_0\,{\rm exp}{\displaystyle \left[\frac{-(t-t_0)^2}{2\tau_0^2}\right]} 
        &~~,~~& t < t_0 \\[12pt]
          \lambda I_0\,{\rm exp}{\displaystyle \left[\frac{-(t-t_0)^2}{8\tau_0^2}\right]} &~~,~~& t > t_0 ~~. \end{array} \right.
          	\label{Qt}
\end{equation}
The volumetric heating rate used in EBTEL is given by $Q(t) = H(t)/L$. This asymmetric profile differs from 
the symmetric form used by \citet{Qiu2012} and \citet{Liu2013}, owing to its long decay time.
This long decay is partly due to gradual cooling of a flare loop leading to elongated transition-region
emission at the flare foot-points \citep{Qiu2013}, but continuous heating may also contribute to it. 
Here we empirically take the heating timescale after the peak $t_0$ to be twice the rise time $\tau_0$. 
The free parameter $\lambda$ converts the observed count rate of UV light curve, in DN per second 
per pixel, to a heating rate in units of erg s$^{-1}$ cm$^{-2}$. This parameter is the same for all half loops, and is found by 
matching high temperature emission in AIA 131 channel between the model and the observations.

In the EBTEL model we adopt the prescribed parameters that scale the mean temperature of the loop and 
temperature at the base of the loop to the peak temperature \citep{Klimchuk2008}. Conduction flux is calculated 
with classical Spitzer-H\"arm conductivity. The 0d results with these prescribed parameters have been bench-marked with 
the 1d hydrodynamic simulations \citep{Klimchuk2008}, and variation of these numbers has rather insignificant impact 
on the model synthetic results for thousands of loops \citep{Liu2013}. We also use another free parameter to 
characterize the loss term through the transition region. Instead of computing this loss term using an equilibrium
solution \citep{Cargill2012a}, we scale this term as being proportional to the mean coronal pressure $\langle P\rangle$ by a scaling 
constant $\eta$. Such proportionality is observed in the decay phase of solar and stellar flares as
well as predicted in coronal heating models \citep[][ and references therein]{Hawley1992, Qiu2013}.  We set $\eta$ by matching  
low-temperature emission in AIA 171 channel between model and observation. Experiments have 
shown that the synthetic plasma emissions at high and low temperatures are independently sensitive to 
the two parameters, $\lambda$ and $\eta$, respectively \citep{Qiu2012}. 

\subsection{Global Evolution Pattern}

\begin{figure}
\epsscale{1.0}
\plotone{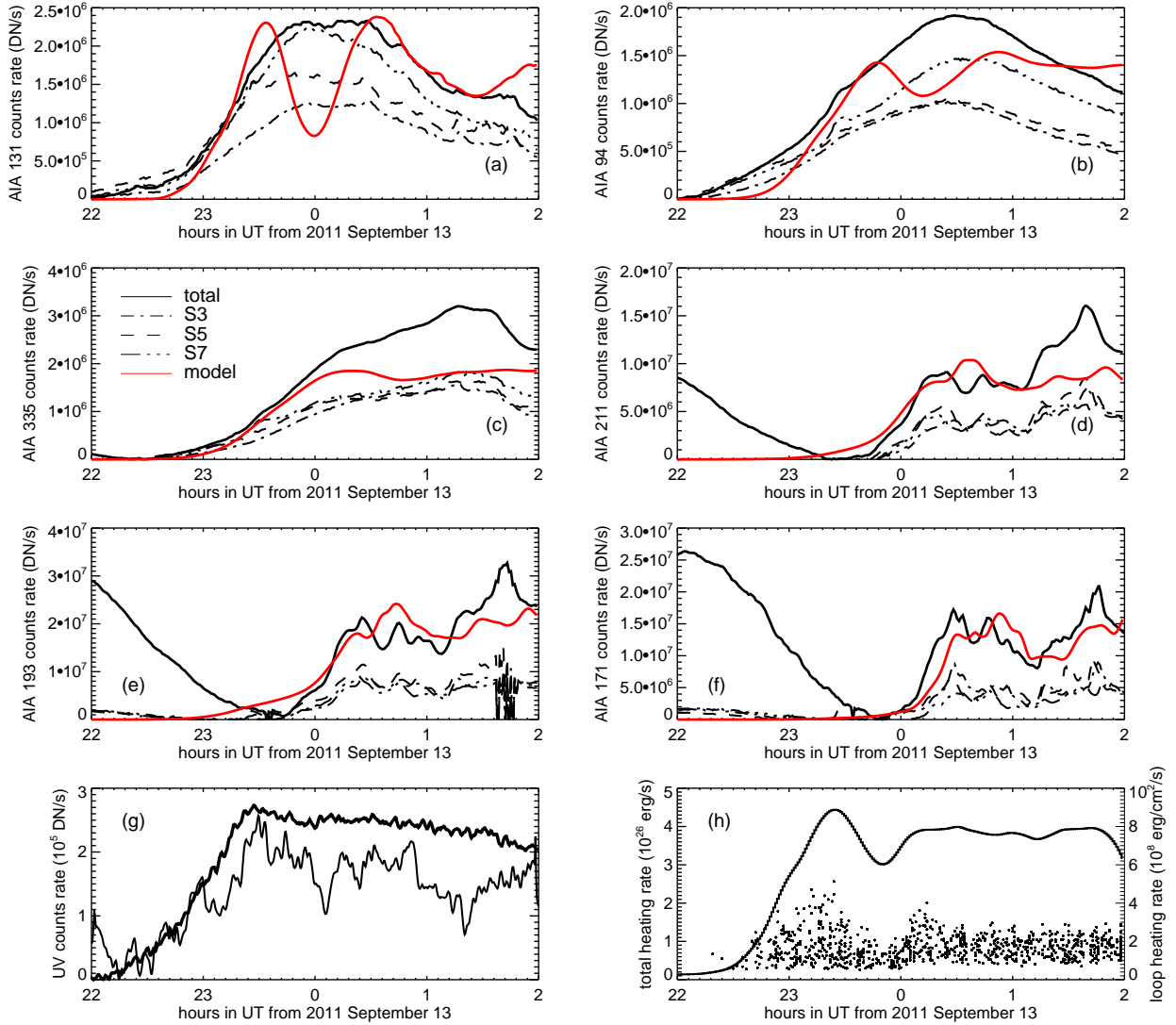}
\caption{(a) - (f): observed (black) and synthetic (red) EUV emissions of the flare in 6 AIA bands. Solid black curves show the
total emission in the active region; also shown in black is the total emission of the loops along slit S3, S5, and S7, 
respectively. For display, the observed emission along a slit is multiplied by a factor of 50. (g): AIA observed total UV
1600 emission (thick black) in flaring pixels in the rectangle marked in Figure~\ref{intensitygram}, in comparison with one-half
of the total UV 1600 emission from the entire active region (thin black).
(h): total heating rate in all flare loops and peak heating rates in each heating event (black dots), derived from UV 
lightcurves of flare ribbon pixels.} \label{obsmdl}
\end{figure}

Figure~\ref{obsmdl} shows the total synthetic emission (red) from all 12,500 full loops (found by doubling the emission from 
the half loops) in six AIA EUV bandpasses. These are plotted atop the total counts rate from observations (black) with no 
scaling or normalization applied. It is seen that the overall timescale and emission levels are reproduced rather satisfactorily by 
the UFC method.  The agreement in all six bands is achieved through the adjustment of only two free parameters ($\lambda$ and $\eta$) 
with reference to only two bands, as described above.  Notable discrepancies include the dip in the total synthetic 
emission in 131 and 94 bands around the flare peak time at 0~UT, not present in the observations, and the 
insufficient emission levels (a deficit by a factor of 1.5) in 335 and 211 bands after 1 UT. 
The dip in the synthetic 131 emission is coincident with a drop in the observed total UV 1600 emission 
(see Figure~\ref{overview}, and bottom left panel of Figure~\ref{obsmdl}), 
indicating a gap of heating events around this time. Nevertheless, the decrease in the UV emission is rather mild compared with the synthetic 131 emission,
which is only one third of the observed emission. It is likely that our model does not provide
enough heating during the 30 min between 11:30 and 0:00~UT. Heating of each of these loops might proceed for a longer time 
to generate more total emission toward 0:00~UT. This will be further discussed in the next section.

For further comparison, the summed emissions from three of the artificial slits, 
S3, S5, and S7, are plotted in each figure after multiplication by an arbitrary factor.  
These plots show that the 0d model produces total light curves whose temporal profile 
generally agrees with those from different portions of the actual loops. 

The lower left panel shows the UV count rate summed over all 12,500 flaring pixels (thick black), 
in comparison with the overplotted UV light curve of the entire active region (thin black), 
the latter being multiplied by a factor of 0.5. These plots show that UV emission from the 
narrow positive ribbon accounts for over 50\% of enhanced total UV emission in the active region, 
and the two light curves exhibit similar trend. 
It would therefore seem that our analysis captured most of the heating events, provided that they produce enhanced
UV emission at the flare foot-points.

The amplitudes of the individual heating events, $\lambda I_0$, are plotted as black dots in the bottom
right panel. The peak amplitude is 5.1$\times 10^{8}$ erg s$^{-1}$ cm$^{-2}$, and the median is 
$\lambda I_0=1.8\times 10^8$ erg s$^{-1}$ cm$^{-2}$. The integral of $H(t)$ for all loops gives the total 
heating rate plotted in the solid line. This flare is powered by heating $\simeq 4.0\times 10^{26}$ erg s$^{-1}$, delivering a modest total energy 
$4.3\times 10^{30}$ ergs over about four hours. Of this total, the model predicts that 1.1$\times 10^{30}$ ergs 
are radiated from the corona. For comparison, the total radiative loss derived from the GOES two-channel diagnostics 
using the standard SSWIDL package is 1.6$\times 10^{30}$ ergs (not shown here) -- note that this value is an over-estimate 
of coronal radiation for the studied event since the GOES emission is contaminated by the other flare occurring at the same time.

For a more revealing model-to-data comparison, we synthesize time-distance stack plots like those shown in the three left columns of 
Figure~\ref{intensitygram}. A synthetic stack plot is produced by summing all synthetic emission from loops whose 
UV pixels fall within the same slit-position. There are on average 12 such loops in a slit position, but the central 
region 100\arcsec\ -- 180\arcsec\ has more. These synthetic stack plots are arranged along the right column of 
Figure\  \ref{intensitygram}, beside the three observed stack plots from the same bandpass.  
Since EBTEL is not spatially-resolved, it produces only one stack plot for each bandpass.  

The synthetic stack plots share some characteristics with the observed versions in that the general 
tempo-spatial sequence of energy release, heating, and cooling along the PIL is reproduced. However, large 
discrepancies are also present. The 
synthetic emission exhibits a much narrower and nearly bi-modal distribution, in contrast to the observed 
smooth and broad emission pattern. For example, the synthetic plots show enhanced brightness in 131 emission at 
around 160\arcsec - 170\arcsec along the slit and no heating beyond, whereas the observed emission extends further and more
smoothly. One source of such discrepancy lies in the assumption that the spatial coordinate 
of the UV footpoints matches the coordinate where the loop crosses a slit. Observations show that loops formed later 
are more sheared than earlier-formed loops; as a result, the observed arcade extends further south than does the ribbon and, 
therefore, the synthetic arcade emission constructed using the ribbon position. The discrepancy in the
temporal distribution is harder to explain. Even in the earlier phase when the ribbon
positions nearly match the loop positions, at each slit position, individual loop pixels remain bright much longer in reality 
than they do in the model. This leads to the deficit of the total synthetic emission seen in the light curves 
in Figure~\ref{obsmdl}.

\subsection{Single Loop Statistics}

The nature of the discrepancy between model and observation can be further explored using the
statistics of individual loop pixels.  The peak times of synthetic EUV emission, shown with black dots in
Figure~\ref{stat}b, fall mostly within the range of the observed peaks (colored curves).
Model-observation agreement is best in the high temperature 131~\AA\ band. This agreement suggests that the tempo-spatial 
distribution of heating events is reasonably captured by the model. The durations and delays of 
these peaks, whose histograms are plotted in red in Figure~\ref{stat}c and d, show a more pronounced 
discrepancy with observation (black).  Model loops have substantially shorter duration in the hotter 
bands (131 and 94) than observed; the deficit is about 30 min.\  (see Table 1 for values).  This 
failing leads to the narrower bands noted in the stack plots of Figure~\ref{intensitygram}.
This duration discrepancy is absent in the cooler bands.  The inter-band delays (bottom rows) show a similar pattern: 
hotter bands cool more slowly in observations while the cooler bands appear to agree.

One possible explanation of the discrepancy is that our UV-enhancement criteria captures only the strongest heating events, 
and may miss a significant contribution from weaker events.  To test this hypothesis, we
plot, with dashes in Figure~\ref{stat}, histograms from the subset of the 15\% brightest pixels (based on 131\AA\ emission). 
There is no evident difference in the distributions of the very brightest pixels and the entire sample.  This is equally 
true of the model (red) and the observations (black). 

\begin{figure}
\epsscale{1.0}
\plotone{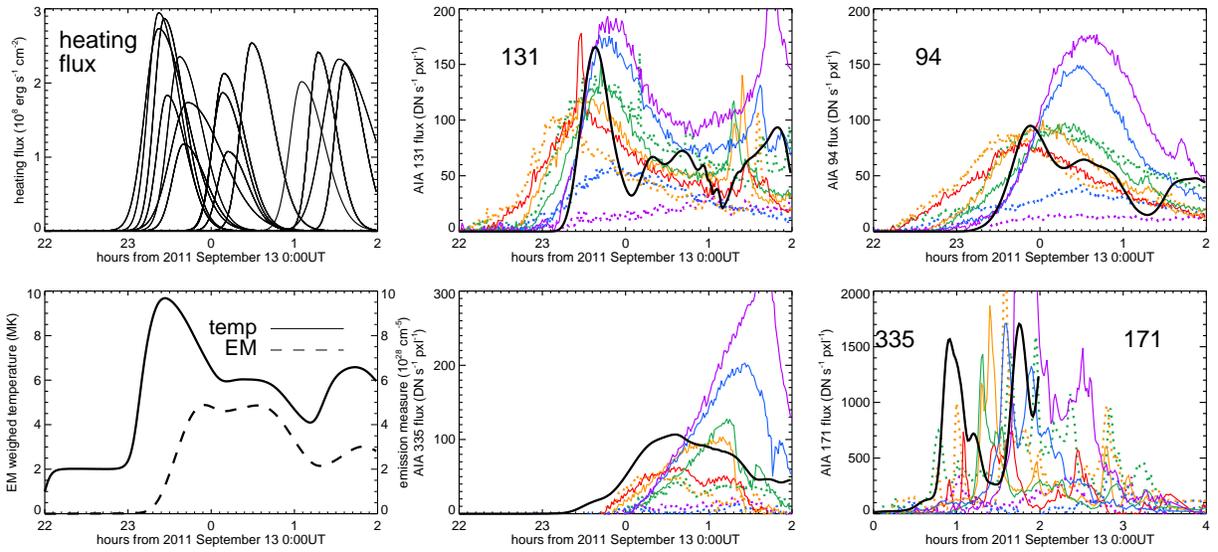}
\caption{Upper-left: heating rate profiles inferred from UV light curves at a slit location across the ribbon. Lower-left: 
EBTEL calculated Emission Measure (EM) and EM weighed temperature of multiple loops using the inferred heating rates.
Middle and right: synthetic AIA emission (in terms of DN/s) at one loop pixel by multiple loops stacked on top of 
each other along the line of sight, in comparison with observed AIA emission at one loop pixel in multiple bands. Different
colors and line styles show the observed loop emission in pixels along different slits as in Figure~\ref{stat}b. } \label{onepixel}
\end{figure}

Another expanation is that an observed loop pixel at a slit position does not exactly match the foot-point pixels
of the ribbon at the same slit position as said before, and also loops may expand and entangle in the corona. For these reasons,
emission at a loop pixel is possibly contributed from a few heating events at different times. 
Figure~\ref{onepixel} explores this possibility with an example of the observed and synthetic 
emission from one location along the slits. More than a dozen heating events, found in UV pixels, 
were assigned to this one slit location. Plotted in the top left panel are the heating flux of these events 
spread in 3 hours. Emission measure (EM) as well as the EM-weighed temperature of the resulting model 
loops are plotted in the bottom left panel. The synthetic emission at one slit pixel is then integrated along
the line of sight and presented in the other panels in comparison to the observed AIA emission. The model curves are 
plotted in DN per second per pixel with no scaling or normalization. 
It is evident that the introduction of multiple heating events spread over 3 hours produces synthetic 
emission for as long as the observed examples. It does not, however, produce in the synthetic emission 
the wide envelop (long duration) actually observed in the high temperature emission --- particularly the 131 emission.

\subsection{Lessons from the UFC}

The standard UFC model, applied above, leads us to the following understanding:

\begin{enumerate}
\item the UFC method with the 0d multiple loop model is able to reproduce the observed global 
evolution pattern to the first order (Figures 2, 3b, 4);
\item superposition of multiple heating events also produces the overall evolution timescale
of EUV emission at some pixels, consistent with observations (Figure 5);
\item however, the majority of individual loop pixels tend to evolve more slowly from 10 - 3~MK, and the duration of 
emission in 131 is much longer (by 30 minutes) than model results (Figure 3c, 3d, Figure 5);
\item on the other hand, these same pixels are observed to evolve very quickly at the low temperature below 3MK;
in particular, 171 emission exhibits multiple bursts each having a duration of 5-10 min. 
Modeled and observed duration of the 171 emission is in good agreement (Figure 3c);
\item the foot-point UV light curve exhibits a long decay time of 10 - 60 minutes, 
maybe indicative of a long continuous heating at individual ribbon pixels as can be resolved by AIA ($~$1-2\arcsec) (Figure 3d);
\end{enumerate}

\section{Modified UFC modeling}

The agreement with global evolution patterns and discrepancies with individual
loop pixels indicate that the former is primarily determined by the temporal distribution of heating events, and
is apparently captured by the UFC method. The single-pixel discrepancy in duration
of the hotter bands may be attributed to unresolved threads anchored to the same footpoint pixel, 
heated independently at different times throughout the duration of the UV brightening. To test this hypothesis, 
we explore several extensions of the UFC model to account for a single {\em loop} comprising multiple {\em threads}. 

We illustrate the different extensions using a single loop whose
footpoint pixel exhibits a UV rise time $\tau_0 = 7$ min, and peak heating rate
$H_0=\lambda I_0 = 3\times 10^8\ {\rm erg\ s^{-1}\ cm^{-2}}$. We call this loop the reference loop.  
The integral of heating profile eq.\ (\ref{Qt}) yields a total heating of $4.7\times 10^{11}\ {\rm erg\ cm^{-2}}$ 
delivered to this reference loop.  We retain all of these 
observationally-determined parameters, $H_0$, $\tau_0$, and $t_0$, in all of our multi-thread (MT) experiments, and results
of the experiments will be compared with those of the single or one loop model (OL model hereafter).

\subsection{Multi-thread heating with amplitude modulation -- AM model}

In the first MT experiment, we assume new threads are introduced at a 
constant frequency, but energized to different amplitudes. We call this model the Amplitude Modulation (AM) model hereafter.  
The heating profile of an individual thread is given a double Gaussian profile with a short time scale, $\tau_i$,
allowed to vary from 10 to 150 seconds,  
\begin{equation}
H_i(t) = \left\{ \begin{array}{lcl} H_{i0}\,{\rm exp}{\displaystyle \left[\frac{-(t- t_{i0})^2}{2\tau_i^2}\right]}
        &~~,~~& t < t_{i0} \\[12pt]
          H_{i0}\,{\rm exp}{\displaystyle \left[\frac{-(t- t_{i0})^2}{8\tau_i^2}\right]} &~~,~~& t > t_{i0} ~~, \end{array} \right.
                \label{AMI}
\end{equation}
where $t_{i0}$ is the peak time for the $i^{\rm th}$ thread.  These impulsive pulses are spaced at 
regular intervals of $\Delta t_i \equiv t_{(i+1)0}-t_{i0} = k \tau_i$, which is $k$ times the pulse width $\tau_i$.  
The peak amplitudes follow a two-Gaussian envelope same as the one-loop heating profile (Eq.~\ref{Qt}):
\begin{equation}
H_{i0} = \left\{ \begin{array}{lcl} H_m\,{\rm exp}{\displaystyle \left[\frac{-(t_{i0}-t_0)^2}{2\tau_0^2}\right]}
        &~~,~~& t_{i0} < t_0 \\[12pt]
          H_m\,{\rm exp}{\displaystyle \left[\frac{-(t_{i0}-t_0)^2}{8\tau_0^2}\right]} &~~,~~& t_{i0} > t_0 ~~, \end{array} \right.
                \label{AM0}
\end{equation}
where the width of the envelope, $\tau_0$, is fixed by the rise-time of the UV light curve of the footpoint.  

The maximum peak is scaled to the observed peak, $H_m = fH_0$, by a factor $f$ chosen to be in 
the range from 1 to 30. A time integral of the composite heating function for the MT experiment is
\begin{equation}
  \int\, dt\, \sum_i H_i(t) ~=~ 1.5 \sqrt{2\pi}\,\sum_i \tau_i H_{i0}
  ~\simeq~ \frac{(1.5\sqrt{2\pi})^2}{k}\, f\tau_0H_0 ~~.
  \label{Hit}
\end{equation}
The OL heating rate, given by Eq.\ (\ref{Qt}), has an integral of $1.5\sqrt{2\pi}\tau_0\,H_0$.  These integrals must be 
multiplied by the corresponding cross-sectional area to obtain the total heating energy. Each thread presumably has a cross-sectional area
smaller than the cross-sectional area of the reference loop by some factor $\zeta$.  If we set $\zeta = k/(f1.5\sqrt{2\pi})$, 
then the total energy delivered to the MT model will match that delivered to the OL model, for 
any choice of the frequency parameter $k$ and the heating rate parameter $f$. We may therefore consider 
$k$ to be an entirely free parameter, related to the ratio of cross sectional areas $\zeta$. Whereas this 
uncertainty cannot be fixed by current observations, the cross-sectional area itself does not 
have any impact on a 0d or 1d hydrodynamic model. On the other hand, the energy flux per unit area, 
as modulated by the other free parameter $f$, is critical to the model calculated properties
and the resultant synthetic loop emission.

The mean temperature $\langle T\rangle$ and electron density $\langle n_e\rangle$ of each thread computed with the 0d model are then 
used to calculate the synthetic EUV emission at different AIA bandpasses, measured in units of DN per 
second per pixel. Since we assume that these threads comprise the reference loop, they all cross 
the same one AIA pixel on the reference loop, and the total synthetic emission by all threads
at this AIA pixel is computed.

\begin{figure}
\epsscale{0.7}
\plotone{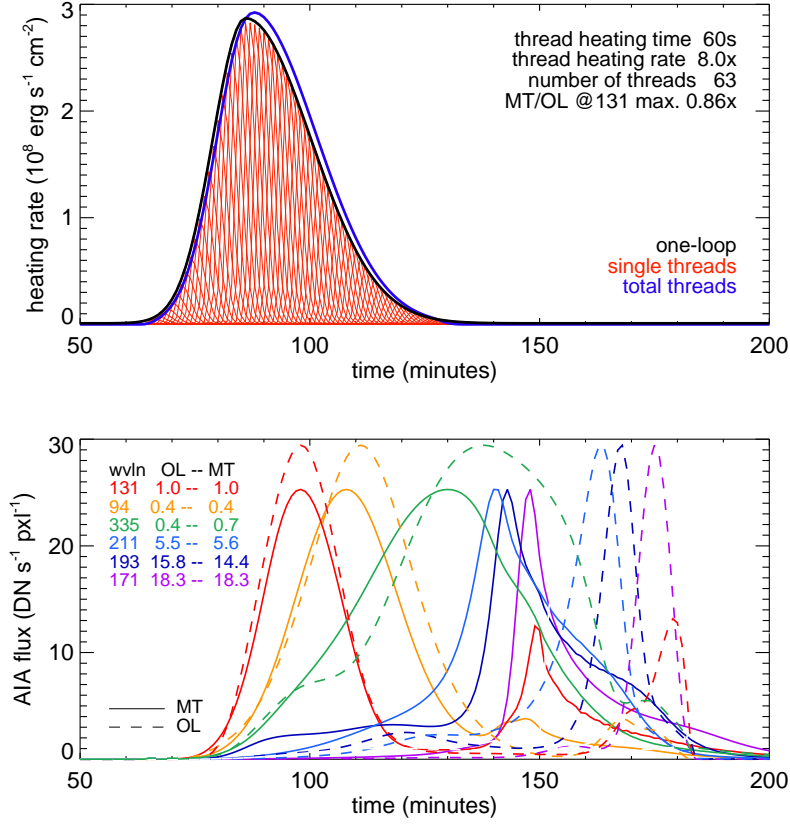}
\caption{An example of multi-thread AM model result in comparison with the one-loop 
model (OL) result. Top: heating flux of the OL model directly inferred from the UV light curve in one AIA pixel (black), 
prescribed heating fluxes of multiple threads (63 threads in this example) in this pixel (red), and total heating flux of the multiple
threads (blue). In this example, the peak heating rate of the multi-thread is 8 times the
one-loop peak heating rate. For better display, heating flux in each thread is re-scaled by a factor of 8.
The heating time in the OL model is 7 min, and the heating time of each thread in the AM model is 60s. The total heating energy
of the threads is the same as in the OL model. Also given is the amount of synthetic peak AIA emission
in 131 band from the AM model, which is 0.86 of the peak AIA emission in the same band from the OL model.
Bottom: comparison of synthetic AIA emissions in multiple channels at one AIA (loop) pixel computed with the multi-thread
AM model (solid) and with the OL model (dashed). Note that the scale on the y-axis shows the synthetic AIA emission in 131 band, and emissions in all other bands are scaled to the 131 emission, with the scaling
factor marked in the figure. The peak synthetic 131 emission by the AM model is 0.86 that of the OL model for
this example.}   \label{constnum}
\end{figure}

An example of the AM experiment is shown in Figure~\ref{constnum}. In this example we have taken $f=8$ 
and $\tau_i = \Delta t_i = 60$~s (i.e., $k = 1$) to give the peak heating rate $H_m = 2.4\times 10^9\ {\rm erg \ s^{-1} \ cm^2}$.  
In the time period of $9\tau_0$, from $t_0 -3\tau_0$ to $t_0 + 6\tau_0$, a total of $N = 9\tau_0/(k\tau_i) = 63$ 
threads were introduced. The heating rates $H_i(t)$ of these threads, and their total heating rate $\sum_i H_i(t)$
are plotted against the heating rate $H_0(t)$ of the reference loop in the top panel.
To match the total energy of the 63 threads with the energy in the reference loop, we 
find $\zeta = 0.03$, or the diameter of each thread is about 17\% that of a loop. Obviously with a longer interval of threads, or 
$k = \Delta t /\tau_i > 1$, the cross-sectional area of the thread will be greater.

The bottom panel of Fig.~\ref{constnum} shows the synthetic AIA emission at different bandpasses by the multi-thread AM model, compared 
with the OL model. With this specific parameter set $f = 8$ and $\tau_i = 60$~s, the synthetic AIA emission in 131 bandpass is comparable with 
the OL model, the peak 131 emission being about 86\% of that by the OL model. The ratios of the peak emissions in other bandpasses
relative to the peak 131 emission are quoted in the bottom panel of the figure, for the AM model and OL model, respectively.
Again, for this set of parameters, EUV emissions in various AIA bandpasses by the AM model are comparable with those by the OL model.
However, the duration of the synthetic 131 emission by the AM model is not longer compared with the OL model, 
and the cooling times between different bands as derived from the AM model are actually shorter than from the OL model. 

The AM experiment has been conducted with many sets of $f$ (ranging from 1 to 30) and $\tau_i$ (from 10 to 150~s). 
These experiments show that duration of the synthetic 131 emission is nearly determined by the timescale of the total heating $\Sigma H_i(t)$, 
and is rather insensitive to properties of the threads such as $\tau_i$ and $H_{i0}$. The OL model produces longer cooling
times between different bandpasses than any of the AM experiment, suggesting that persistent heating in a loop or thread 
is indeed important for the long cooling time.

\subsection{Multi-thread heating with frequency modulation -- FM model}
In the second MT experiment, threads are heated with identical profiles and
amplitude but new threads are introduced at a rate varied to reproduce the observed UV light curve.
This is the Frequency Modulation (FM) model. The thread production rate is given a double Gaussian 
profile resembling the $H_0$ profile
\begin{equation}
{dN\over dt} = \left\{ \begin{array}{lcl}  \dot{N}_m \,{\rm exp}{\displaystyle \left[\frac{-(t_{i0}-t_0)^2}{2\tau_0^2}\right]}
        &~~,~~& t_{i0} < t_0 \\[12pt]
           \dot{N}_m\,{\rm exp}{\displaystyle \left[\frac{-(t_{i0}-t_0)^2}{8\tau_0^2}\right]} &~~,~~& t_{i0} > t_0 ~~, \end{array} \right.
 	\label{dNdt}
\end{equation}
where $\dot{N}_m$ is the peak rate of the thread, and $\tau_0$ is the same rise time used above.
The individual heating profiles are given by eq.\ (\ref{AMI}) with a constant peak rate, $H_{i0} = fH_0$ and 
time scale $\tau_i$ for each thread.  We perform experiments in which $\tau_i$ is set to the values 10, 30, 60, 150~s, 
and where $f$ ranges from 1 to 30, respectively. 

\begin{figure}
\epsscale{0.7}
\plotone{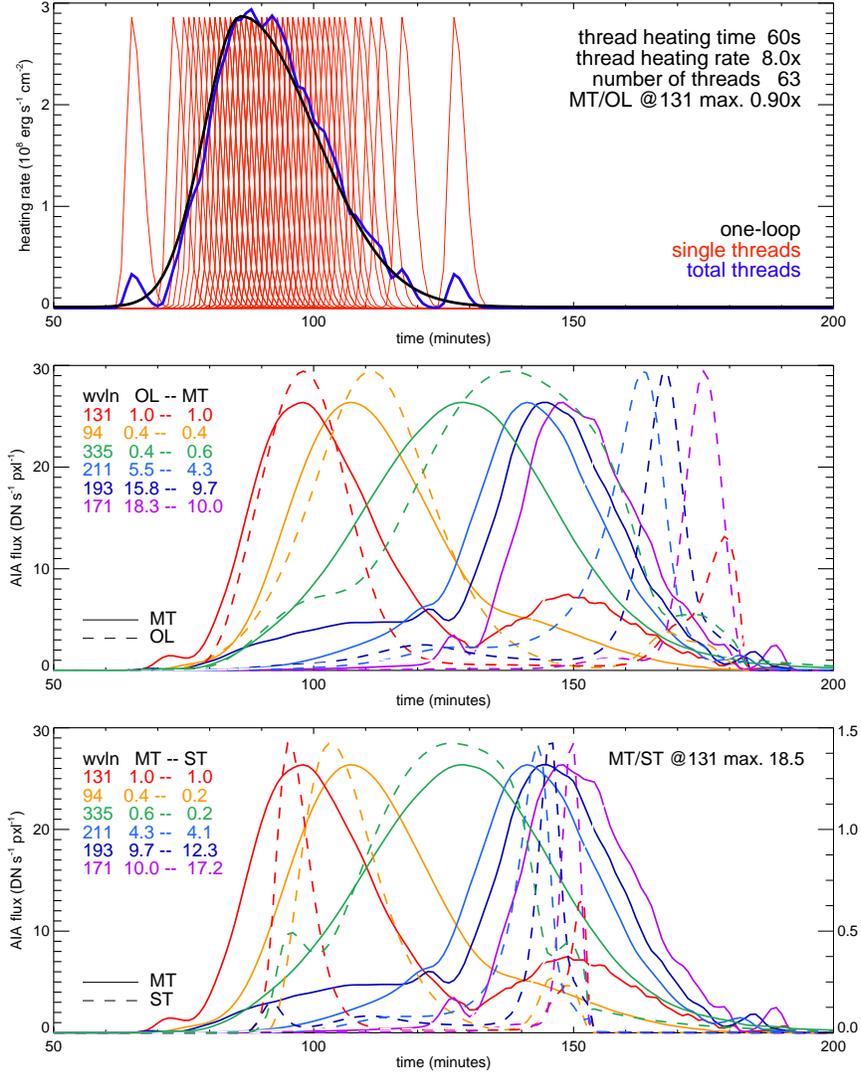}
\caption{An example of multi-thread FM model result in comparison with the OL model result. 
Top: heating flux of the OL model (black), prescribed heating fluxes of multiple threads (63 threads in this example) 
in this pixel (red), and total heating flux of the multiple threads (blue). In this example, the peak heating rate 
of each thread is 8 times the peak heating rate of one-loop. For better display, heating flux in each thread
is re-scaled by a factor of 8. The OL model is the same as in Figure~\ref{constnum}; and the heating time of 
each thread in the FM model is 60s. The total heating energy of the threads is the same as in the OL 
model. The amount of synthetic peak AIA emission in 131 band from the FM model is 0.90 of that from the OL model.
Middle: comparison of synthetic AIA emissions in multiple channels at one AIA (loop) pixel computed with the multi-thread 
FM model (solid) and with the OL model (dashed), same as in Figure~\ref{constnum}. Bottom: comparison of the synthetic AIA emissions in multiple channels at one AIA (loop) 
pixel computed with the multi-thread FM model (solid; scale on the left axis) and the synthetic emissions by one of the threads
(dashed; scale on the right axis). 
Note that the scales show the synthetic AIA emission in 131 band, 
and emissions in all other bands are scaled to the 131 emission, with the scaling factors marked in the figure.} \label{constrat}
\end{figure}

Again, to match the total energy of the threads to that in the reference loop, the cross-sectional
area of each thread as some fraction $\zeta$ of the area of the reference loop is given by
$\zeta = 1/(1.5 \sqrt{2\pi} f\dot{N}_m\tau_i)$. Figure~\ref{constrat} shows an example 
of FM heating with $f = 8$, or $H_{i0} = 2.3\times 10^9\ {\rm erg\ s^{-1}\ cm^{-2}}$, 
and $\tau_i = 60$s. We populate the threads with the peak rate $\dot{N}_m=2.4$~ per min, which again
introduces 63 threads during the period from $t_0 - 3\tau_0$ to $t_0 + 6\tau_0$.
The top panel displays heating profiles of the threads and their total heating rate 
in comparison with the OL heating profile. To match the total energy to the reference loop, 
$\zeta \sim 0.014$, or the thread diameter is 12\% of the loop diameter. Again, $\dot{N}_m$ and therefore $\zeta$ are free parameters 
that can be re-adjusted, with no impact on the 0d and 1d model.

The synthetic AIA emissions at different bandpasses are calculated and displayed in the middle panel. For 
reference, the synthetic AIA emission by one thread is also displayed in the lower panel. In this example 
of 63 threads, the peak 131 emission from the FM model is about 90\% that from the OL model, if both use the 
same amount of heating energy. Compared with the first experiment (AM), the constant amplitude model (FM) is 
capable of producing 131 emission of long duration. This is primarily due to strong heating events in 
the early rise phase as well as the decay phase. The superposition of these early and late threads gives 
rise to longer duration of the 131 emission. Shown in the bottom panel, the duration of the 131 emission by a single thread is 
rather short, so a reasonable number of threads would be needed to produce the smooth 131 emission in the loop
as observed. 

The FM experiment fails, however, to reproduce the long cooling delays. It seems the strong 
impulsive heating raises the density of the plasma quickly, leading to enhanced radiative cooling and therefore 
a much shorter cooling time from 131 through 94, 335, and 171 passbands, than found in the OL model. 
In addition, we also note that, as shown in the bottom panel of the figure, the durations of the synthetic 
131 and 171 emissions by a single thread are nearly comparable; therefore, whereas superposition of emissions by many threads could 
produce the long-duration and smooth emission in the 131 bandpass, it may also lead to long-duration 171 emission, which is not observed.
The OL model, with persistent yet attenuated heating, produces the longest cooling timescales from 10 - 3~MK as well as 
the short duration of 171 emission comparable to observed timescales.  

While none of the foregoing experiments in UFC modeling is completely successful, they do provide insight into 
the ingredients required to successfully reproduce observations.  First of all, strong impulsive heating in the early as well as
decay phase is necessary to produce the long duration 131 emission at 10~MK observed in the SOL2011-09-13T22 flare. 
On the other hand, matching the long cooling delays requires continuous, lower-amplitude heating following the impulsive
phase in the same thread.  This low-amplitude heating nearly balances the radiative and conductive
losses and therefore maintains the same (high) density of the thread; as the thread gradually cools to below 2~MK, radiative 
cooling increases rapidly and the timescale becomes very short generating the short duration 
emission at 1-2 MK. We use this approximate balance to estimate the amplitude of the continuous heating required
in the gradual phase, and
propose that the heating profile in a thread must start with an intensive impulsive heating pulse, followed by a persistent heating 
at the rate one to two orders of magnitude lower, which is gradually attenuated to maintain the observed timescales 
of emission in the temperature above 3~MK. 

\subsection{Single thread with impulsive head and slow tail - ST model}

Based on these analyses, we design the third experiment for a single thread, which is heated impulsively
at the rate of $H_{im} = 10^{9-10} {\rm erg\ cm^{-2}\ s^{-1}}$ for timescale $\tau_{im} = 10 - 150$ s, followed 
by a persistent low-rate heating starting at the maximum rate of $H_{sl} = 0.01 - 10\times 10^8\ {\rm erg\ cm^{-2}\ s^{-1}}$ 
for a timescale of $\tau_{sl} = 5 - 30$~min. The model is called the ST (Slow Tail) model.
The heating profile of this model is composed of several Gaussian components
\begin{equation}
  H(t) ~=~ \left\{ \begin{array}{lcl} H_{im}\, {\rm exp}{\displaystyle \left[\frac{-(t-t_0)^2}{2\tau_{im}^2}\right]} 
        &~~,~~& t < t_0 \\[12pt]
        H_{im}\, {\rm exp}{\displaystyle \left[\frac{-(t-t_0)^2}{2(2\tau_{im})^2}\right]} 
        &~~,~~& t_0 < t < t_1 \\[12pt]
        H_{sl}\, {\rm exp}{\displaystyle \left[\frac{-(t-t_1)^2}{2\tau_{sl}^2}\right]}  &~~,~~& t > t_1 ~~,
        \end{array} \right. 
        	\label{eq:Ht-tail}
\end{equation}
where $t_1 = t_0 + \tau_{im}\sqrt{8\ln(H_{im}/H_{sl})}$, in order that the profile be continuous.
For each set of parameters, the EBTEL model is run, synthetic 
AIA emission by this single thread is computed, and the durations and cooling times are compared with the results 
of the one-loop heating model as well as the case of a single thread with only impulsive heating ($H_{sl} = 0$). 

\begin{figure}
\epsscale{0.7}
\plotone{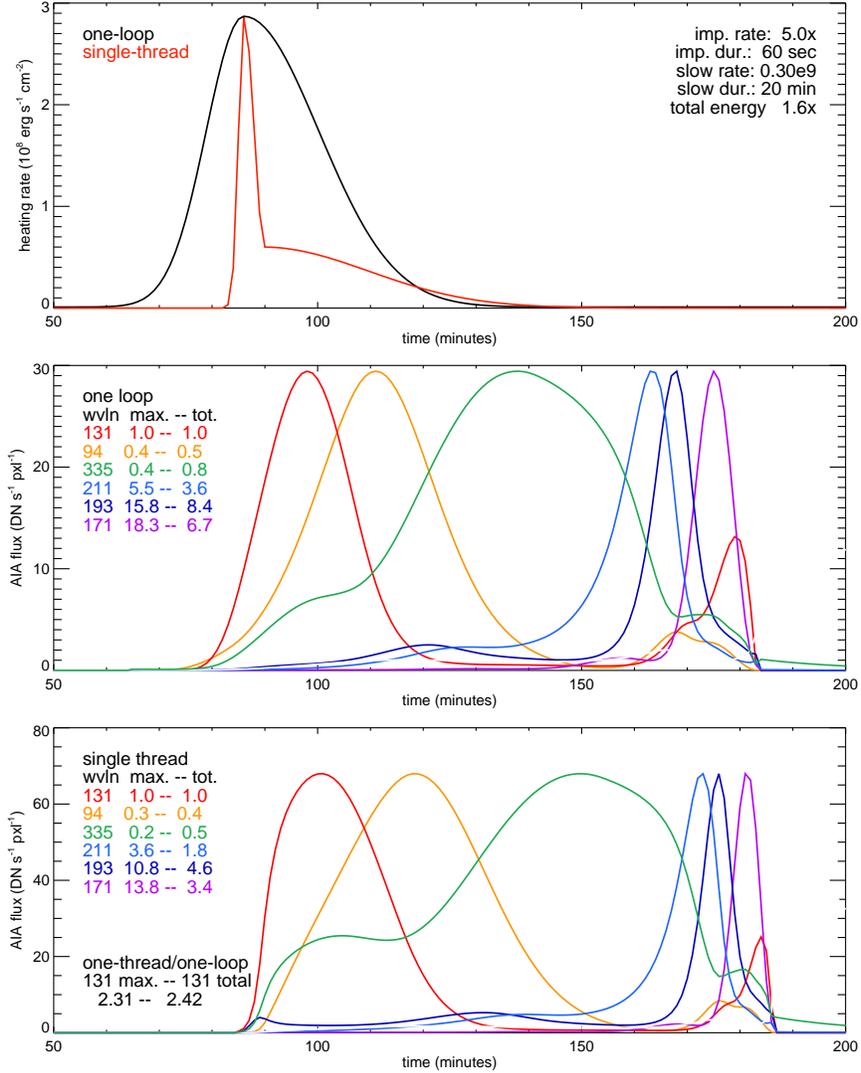}
\caption{An example of single thread model with the slow tail (ST) in comparison with the one-loop model (OL).
Top: heating flux of the OL model, and prescribed heating flux of the single thread in this pixel (red).
The OL model is the same as in Figures~\ref{constnum} and \ref{constrat}. The peak heating rate of the single thread is 5 
times the peak heating rate of one-loop, the heating time of the impulsive phase of the single thread is 60s.
For the slow heating, the peak heating rate is 3$\times 10^8$ erg s$^{-1}$ cm$^{-2}$, and the heating timescale is 20 min.
The total heating energy of this single thread is 1.6 times the total energy in the OL model if both have the same cross-sectional
area. Middle: synthetic AIA emissions in multiple channels at one AIA (loop) pixel computed with the one-loop
OL model, same as in Figure~\ref{constnum}. Bottom: synthetic AIA emissions in multiple channels at one AIA 
(loop) pixel computed with the single-thread ST model. Note that the scale shows the synthetic AIA emission with
the ST model in 131 band, and emissions in all other bands are scaled to the 131 emission, with the scaling
factors marked in the figure. For this example, the peak synthetic 131 emission by the ST model is 2.3 times that of the OL model,
and the total 131 emission by ST is about 2.4 times the total emission by the OL model.
} \label{slowheating}
\end{figure}

Figure~\ref{slowheating} shows the heating profile and the synthetic AIA emissions for one set of parameters
in comparison with the OL model. The heating profile of the thread is plotted in the top panel, in comparison 
with the OL model. The peak impulsive heating rate is $H_{im} = 1.4\times 10^9\ 
{\rm erg\ s^{-1}\ cm^{-2}}$ and timescale $\tau_{im} = 60s$, and the peak slow heating rate is 
$H_{sl} = 3\times 10^8\ {\rm erg\ s^{-1}\ cm^{-2}}$ and timescale $\tau_{sl} = 20$~min. 
If the thread has the same cross-sectional area as the reference loop, then the total energy in the thread 
is 1.6 times that in the reference loop. Again, we may match the total energy of the threads with that of
the reference loop by adjusting the area of a thread relative to the reference loop area and the number (or frequency) 
of threads in a loop. 

The middle and bottom panels show synthetic AIA emissions with the OL and ST models, respectively. 
It is evident that the single thread in this example can produce a long duration (20 min) 131 emission, 
a short duration (4 min) 171 emission, as well as the long cooling time (80 min) between 131 and 171 peaks. 
Importantly, the duration of emission in 131, 94, and 335 bands, as well as the cooling times through these passbands, are substantially longer than 
those in the case of a single thread with only impulsive heating (see the bottom panel of Figure~\ref{constrat}), 
and are also longer than the OL model. On the other hand, duration of the low-temperature emission (211, 193, and 171) 
is rather short and comparable with observed timescales. 

With the timescales generated in the ST model, if several such threads spread out 
with a time interval of 15-20 min, it may produce a smooth 131 emission with the 
duration comparable to the observed timescale of 60 min, as well as multiple, well-separated, 
and short-duration 171 emission peaks at one loop pixel as indicated by observations. 

We also note that, the total AIA emission in 131 band by such a thread is about 1.5 times that 
by the OL model, if both using the same amount of heating energy. The ratios of the total emission in other 
bandpasses relative to the total emission in 131 band are quoted in the middle and bottom panels.
These ratios vary by within a factor of 2 if we compare the ST model with the OL model, suggesting
that redistributing heating energy as specified by the new heating profile will not significantly change
the total EUV emissions from all loops. This heating energy redistribution, however, does signicantly change the timescales of the EUV emissions
in a single loop as described above.

\begin{figure}
\epsscale{1.0}
\plotone{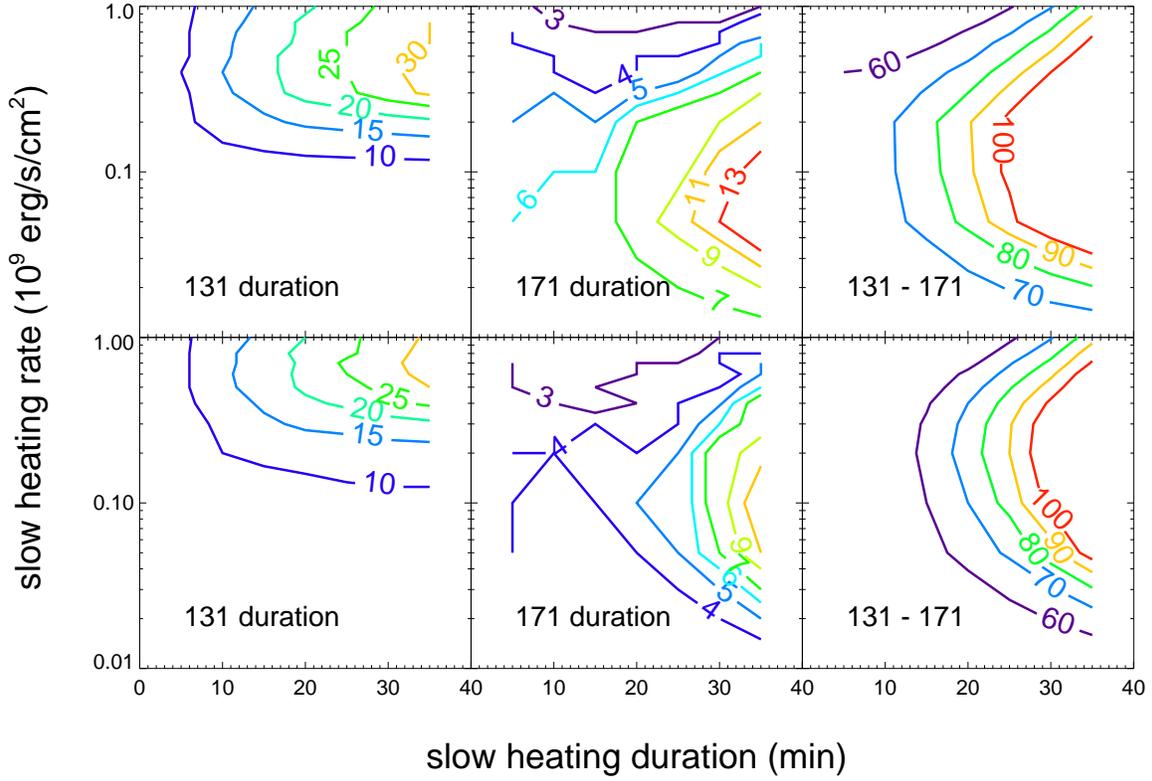}
\caption{Durations and cooling times derived from synthetic AIA emissions in multiple channels using the ST model
with different parameter sets characterizing the heating rate and timescale of the slow-heating in the single thread.
Top: experiments with impulsive heating rate $H_{im} = 1.4 \times 10^9\ {\rm erg\ s^{-1}\ cm^{-2}}$ and duration $\tau_{im} = 60$s. 
Bottom: experiments with $H_{im} = 8.6 \times 10^9\ {\rm erg\ s^{-1}\ cm^{-2}}$ and duration $\tau_{im} = 10$~s.  
} \label{slowstat}
\end{figure}

To explore the range of slow-heating parameters that are able to produce the observed timescales, 
Figure~\ref{slowstat} shows the durations in 131 and 171 bands and cooling time between 131 and 171 emissions
computed with varying parameters of $H_{sl}$ and $\tau_{sl}$. The top panels show the case with an impulsive heating
rate $H_{im} = 1.4 \times 10^9\ {\rm erg\ s^{-1}\ cm^{-2}}$ and duration $\tau_{im} = 60$~s, and the bottom panels
show a case with $H_{im} = 8.6 \times 10^9\ {\rm erg\ s^{-1}\ cm^{-2}}$ and duration $\tau_{im} = 10$~s.
Comparison between the two cases suggests that the duration and cooling time variations of the synthetic AIA emissions
are generally not sensitive to parameters of the impulsive heating, as long as the total energy in the pulse
is comparable; instead, they much depend on parameters of the slow heating. The figures suggest that slow 
heating with the rate $H_{sl} =  2 - 6 \times 10^8\ {\rm erg\ s^{-1}\ cm^{-2}}$ and timescale $\tau_{sl} \ge 20$ min is 
in favor of producing long duration 131 emission ($>$ 20 min), short duration 171 emission ($<$ 8 min), as well as long cooling 
time ($>$ 80 min) between 131 and 171 bands.

\begin{figure}
\epsscale{1.0}
\plotone{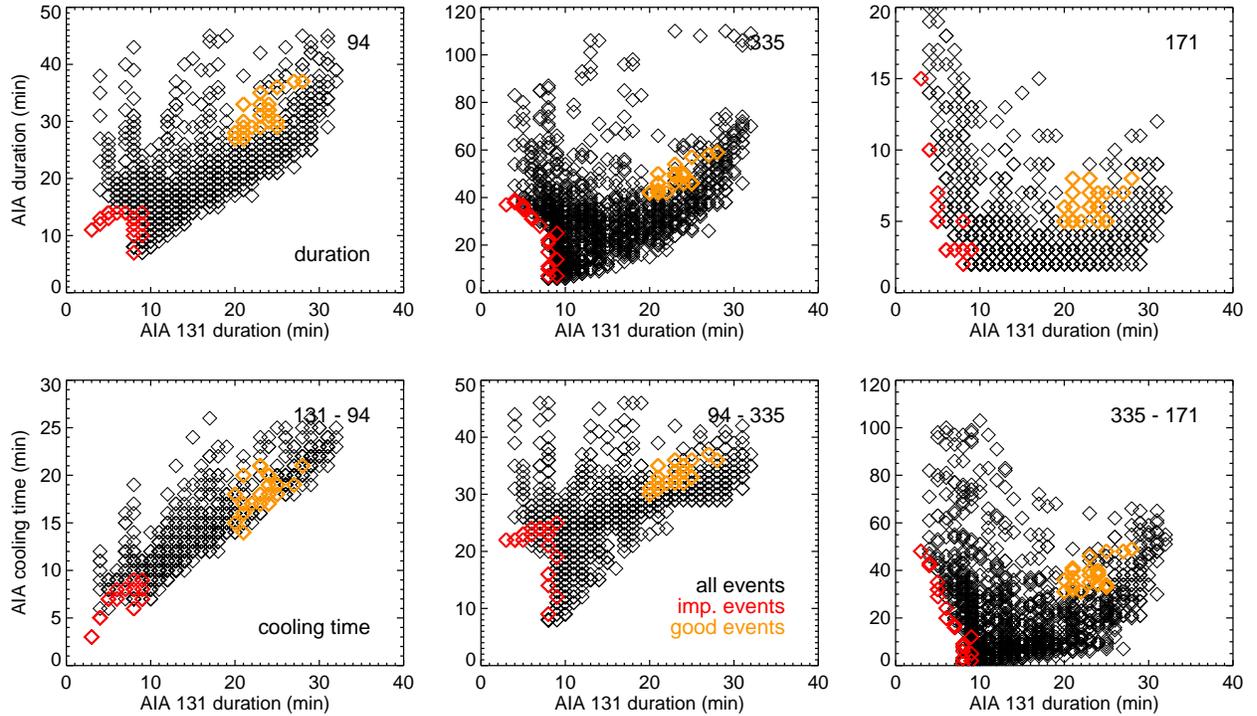}
\caption{Durations and cooling times of synthetic AIA emissions in multiple channels as functions of the synthetic
131 duration using the ST model with multiple parameter sets characterizing the heating rate and timescale of the impulsive
and slow heating in the single thread. Black symbols denote all data points; red symbols show experiments
with only impulsive heating ($I_{sl} = 0$), and orange symbols show experiments yielding desired durations and cooling times
comparable with observations (see text).} \label{slowtrend}
\end{figure}

Figure~\ref{slowtrend} further examines durations and cooling times of AIA emissions in multiple bands as functions
of the 131 duration for 1,400 runs (black symbols) covering a large parameter space: $H_{im} = 1 - 9\times 10^{9}\ {\rm s^{-1}\ cm^{-2}}$,
$\tau_{im} = 10 - 150$~s, $H_{sl} = 0.05 - 1\times 10^{9}\ {\rm s^{-1}\ cm^{-2}}$, and $\tau_{sl} = 15 - 35\ {\rm min}$. 
It is seen that, in general, when duration of the synthetic 131 emission increases, the duration of 94 emission, 
and cooling time from 131 to 94 bands, and then from 94 to 335 bands, also increase. These trends are favorable
to match with observations. On the other hand, the duration of low-temperature emissions at 335 and 171 bands and 
the cooling time between 335 and 171 bands do not grow with increased 131 duration, which is again favorable to 
explain the observed timescales in these bands. Therefore, the experiments show that, by adding a persistent low-rate 
heating component following the intense impulsive heating in the thread, the observed long duration high-temperature 
emissions as well as the long cooling time can be possibly reproduced. In comparison, the data points in red mark the 
cases with only an impulsive heating component; and these heating events produce emissions in all bands for very short durations and 
with very short cooling times as well. 

Finally, data points marked in orange denote the ``good" cases with desired long duration
131 emission ($>20$ min), short duration 171 emission ($<$ 8 min), long cooling time between 131 and 171 bands
($>$ 80 min), and a reasonable ratio of the total 171 emission to the total 131 emission. 
This subset of events selected out of the total of 1,400 runs have a slow-heating rate $H_{sl} = 
2 - 4\times 10^8\ {\rm erg\ s^{-1}\ cm^{-2}}$, and timescale $\tau_{sl} = 20 - 30$ min. 
In these events, the energy in the slow-heating phase is 2-3 times that in the impulsive-heating phase. The 
total heating energy per unit area ranges between 0.4 to 1.0$\times 10^{12}\ {\rm erg\ cm^{-2}}$, which is about 0.8 to 2.2 
times the total heating energy used in the OL model. The total emission in 131 passband increases almost linearly with 
the total heating energy; such linear scaling allows for adjustment of multi-thread parameters, such as the frequency
and cross-sectional area of the thread, to match with the observed amount of total emission. The frequency of the threads
at each loop pixel may be inferred from the multiple 171 peaks, leaving only the cross-section area of the thread
as a free parameter for the modified global modeling. When the modified heating rates
are applied to model all flare loops, refinement of other model parameters including $\lambda$ and $\eta$ may 
also be necessary to match with the observed total emission by the flare. We expect that the adjustment would be minor, 
and defer such modified global modeling to future work to limit the scope of this paper.

\section{One-dimensional Modeling}

The foregoing has used the zero-dimensional EBTEL model to deduce a heating profile which produces light curves sharing 
the properties inferred from observation.  We explore its more general applicability by using the same time-profile 
in a one-dimensional loop-dynamics model.

The one-dimensional run begins with a loop in equilibrium of total coronal length $L=114.6$ Mm.  The equilibrium is maintained by 
uniform coronal heating, $Q_{bk}=10^{-4}\,{\rm erg\,s^{-1}\,cm^{-3}}$, which maintains its apex at $T_{\rm 0,max}=1.3$ MK.  The 
minimum coronal density turns out to be $n_e=3.3\times10^{8}\,{\rm cm}^{-3}$.  We append to each end of the loop, simple isothermal 
chromospheres of $T_{\rm 0,min}=10,000$ K, 4 Mm deep. These include almost eight full scale heights of gravitational stratification, 
over which the electron density rises from $4\times10^{10}\, {\rm cm}^{-3}$, at the top, to $10^{14}\,{\rm cm}^{-3}$ at the lower 
boundary. The chromospheres are intended only as mass-reservoirs, and are treated using coronal dynamics, assuming full ionization 
\citep{Longcope2015,Longcope2014}

The dynamical evolution of the one-dimensional loop is solved using the PREFT code \citep[standing for Post-Reconnection Evolution of a Flux Tube, ][]
{Longcope2015}.  While it is designed to model retracting loops, we run it here with a loop fixed and straight. PREFT includes optically 
thin radiative losses taken from CHIANTI 7.0 \citep{Dere1997, Landi2012}, classical Spitzer-Ha\"rm thermal conductivity, and shear viscosity.  
The initial equilibrium is subjected to an {\em ad hoc} heating source with a tent-profile in space, centered at the apex and extending 
over the central half of the loop. The temporal profile is taken from eq.\ (\ref{eq:Ht-tail}) with
$t_0=120$~s, $\tau_{im} = 60$~s, $\tau_{sl}=1200$ sec, $H_{im}=1.4\times10^{9}\,{\rm erg\,s^{-1}\,cm^{-2}}$, and
$H_{sl}=0.3\times10^{9}\,{\rm erg\,s^{-1}\,cm^{-2}}$.

The heating drives the apex temperature to a maximum value of $T=19.6$ MK by $t=115$~s, just before the heating peaks at 
$t=t_0=120$ s (see eq.\ [\ref{eq:Ht-tail}]).
The heat is conducted to the chromosphere where it drives upward evaporation as well as downward {\em chromospheric condensation}.

The condensation takes the form of a front propagating downward with flow speeds starting at $\pm 70$ km s$^{-1}$, but falling 
steadily to $5$ km s$^{-1}$ by $t=125$~s.  At the same time, evaporation flows at $v\simeq\pm 500$ km s$^{-1}$, form isothermal 
shocks of Mach number about 1.8.  These shocks collide at the loop top around $t=125$ s, creating a reflection which returns 
to the chromosphere at $t=280$ s.  After the reflected waves have returned to the chromosphere, the pressure throughout 
the corona has become uniform to within $\sim5$ \% ($P =25\pm1\,{\rm erg\ cm^{-3}}$). Subsequent evolution consists of 
quasi-static cooling with very small flows and pressure gradients.

The coronal density first rises at the evaporative shocks, by nearly an order of magnitude to
$n_e\simeq2\times10^{9}\,{\rm cm^{-3}}$.  Upon their reflection the apex density rises steadily at a rate $\dot{n}_e\simeq10^{7}\,
{\rm cm^{-3} s^{-1}}$ until peaking at $n_e=7\times10^{9} \ {\rm cm^{-3}} $ at $t=370$ s.  At this point, the corona has reached an equilibrium with 
uniform pressure; equilibrium density is $n_e\simeq10^{10} \ {\rm cm^{-3}} $ at the top of the chromosphere.

Once the one-dimensional model has reached a mechanical equilibrium, at about $t\simeq400$~s, we expect its evolution to be well 
represented by a zero-dimensional model such as EBTEL. It is during the earlier phase, in which density is enhanced through 
supersonic evaporation, that the two methodologies may depart most significantly. It is during this phase that the total mass of 
evaporated material, upon subsequent evolution critically depends, is set.

To make direct comparisons with the 0d runs we follow the definitions used in the formulation of EBTEL.  The corona is defined as the 
region bounded by the points of maximum and minimum conductive heat flux. This region contracts to $L\simeq40$ Mm during the 
evaporation phase, as competing upflows steepen the downward heat flux gradients. During the reflection of the evaporation shocks 
the corona expands back to $L\simeq 120$ Mm.  The coronal pressure and density are found by averaging over the coronal region, so defined.  
The mean density remains above the apex value by 1--3$\times 10^{9}\,{\rm cm^{-3}}$, peaking at 
$\langle n_e \rangle =9\times10^{9}\,{\rm cm^{-3}}$ at $t=463$ s. The 0d temperature (from a ratio of mean pressure to mean density) remains 
below the apex value by 0.5--3 MK, and peaks at the same time as the apex temperature at $t=115$ s, but 1 MK lower, $\langle T \rangle =18.6$ MK.

\begin{figure}
\epsscale{1.0}
\plotone{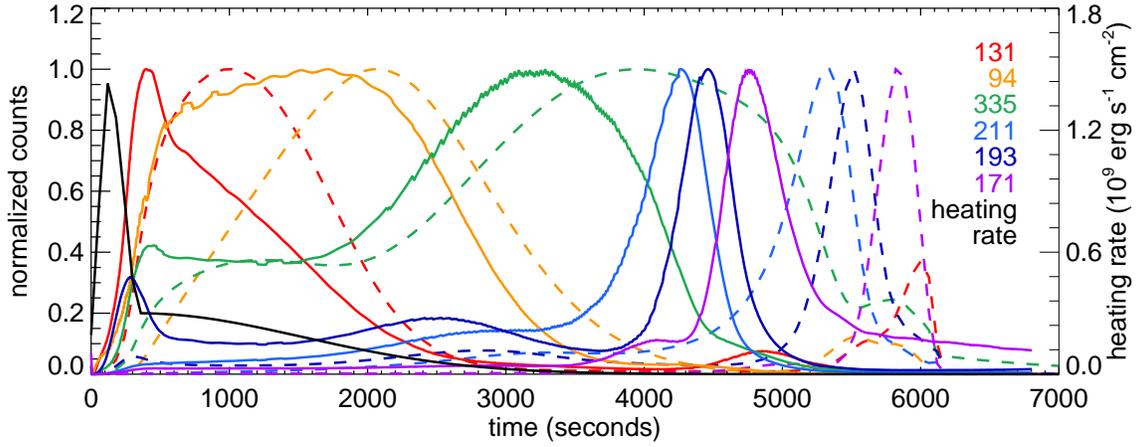}
\caption{Synthetic EUV emissions in multiple AIA passbands produced by the 1d PREFT model (solid) and by the
0d EBTEL model (dashed) using the same heating profile with a prolonged slow-heating tail (black).} \label{onedloops}
\end{figure}

Synthetic light curves are the most significant point of comparison between one-dimensional and zero-dimensional loop models. Figure~\ref{onedloops} 
compares the AIA-observed light curves synthesized from one-dimensional PREFT runs and zero-dimensional EBTEL runs, both with the same heating profiles.
Apart from the first few minutes, light curves from the two models show comparable trends of evolution, namely the long duration
emission in 131, 94, and 335 bands, short duration emission in 211, 193, and 171 bands, and significant delays between high-temperature 
and low-temperature passbands. These features cannot be produced, in either 0d or 1d model, with only an impulsive heating pulse. 
The 1d model therefore qualitatively confirms the general results from the 0d experiments, that the prolonged,
low-rate heating is able to produce observed timescales of plasma evolution. 

The PREFT run does, however, exhibit some notable differences from EBTEL runs with similar heating.  The conduction-driven 
evaporation of PREFT delivers less material to the corona than does the EBTEL. Since 
the conductive cooling time scales positively with coronal density, the PREFT run cools earlier than the EBTEL run. 
In the 0d EBTEL model, evaporation (as well as coronal condensation later on) is computed as the difference 
between conduction from corona and radiation loss in the transition region, the latter term being set to 
be proportional to coronal pressure with a scaling constant $\eta$ in this study (see Section 3). In this sense, 
evaporation in the EBTEL model is parameterized. Furthermore, \citet{Qiu2012} have shown that, in the decay 
phase when impulsive heating has finished, plasma evolution is sensitive to this scaling constant $\eta$ that 
can be adjusted to produce the appropriate timescale as well as the amount of low-temperature emissions.
Physical mechanisms, likely related to the lower-atmosphere dynamics, justifying selection of this parameter 
should be further explored in the ensuing study to improve modeling of flare evolution.

As superior to the 0d model, the 1d model illustrates more accurately plasma evolution in the initial impulsive heating phase when
plasmas are far from equilibrium -- such equilibrium has always been assumed in the 0d model. The 1d model also 
generates along-the-loop plasma properties. These allow comparison with multi-spectral observations
along individual flare loops, which will be reported in the next study.

\section{Conclusions and Discussions}
\subsection{Summary of Results}
In this study, we analyze a long duration flare composed of an arcade of flare loops formed sequentially
along the magnetic polarity inversion line. Using the UFC method \citep{Qiu2012, Liu2013}, we infer heating rates
of thousands of flare loops from the UV light curves at the flare foot-points, and model the flare total 
emission with the 0d EBTEL code. It is shown that the 0d multi-loop model can reproduce the global evolution pattern
of the total EUV emissions, suggesting that the UFC method appropriately captures the distribution of heating events
throughout the flare. However, observations at single loop pixels show long duration EUV emission at high 
temperatures of 6 -- 10 MK, long cooling time from 10 to 3 MK, and very short duration of EUV emission at 1 --
2 MK more than an hour later; all of these signatures at one pixel cannot be produced simultaneously by the model. 

We have then explored the popular thinking that each {\em loop}, assumed to anchor at a UV pixel, is composed of 
unresolved sub-structure, or multiple {\em threads}, heated at different times. Multi-thread models with either 
frequency or amplitude modulation of {\em impulsive} heating events, however, fail to produce the 
observed timescales in loop pixels. Instead, we have found that a heating profile consisting of two parts, an intense 
impulsive heating followed by a persistent low-rate heating, can produce long duration emission at 131 and 94 
passbands, long cooling times, as well as short duration 171 emission. It is estimated that, for each observed loop pixel
in this flare, superposition of a few such heating events (or {\em threads}) with an interval of 10 - 20 min may produce the 
observed timescales at one loop pixel; each of these threads can be heated impulsively at the rate of order $10^{9-10}\ 
{\rm erg\ s^{-1}\ cm^{-2}}$ and timescale $10 - 60$~s, and then gradually with the rate of a few times 
$10^8\ {\rm erg\ s^{-1}\ cm^{-2}}$ that is attenuated in 20 -- 30 min.

\subsection{Discussions}
\subsubsection{Can we see the slow-heating?}
Our experiments have demonstrated that a heating profile consisting of an impulsive component followed by a gradual component
is capable of producing the observed timescales at single loop pixels. However, the observed UV light curves at flare 
ribbons, from which we infer heating rates, do not exhibit the profile similar to the heating profile given in 
Figure~\ref{slowheating}. Can we, therefore, justify the practice of inferring heating rates from the 
foot-point UV light curves? We consider that the observed smooth and long duration UV light curve, 
which does not show the transition between the spiky signature indicative of the impulsive heating 
component and the gradual slow-heating phase, may be caused by the AIA not resolving the threads at
sub-arcsecond scales. Furthermore, the impulsive heating rapidly raises the coronal pressure 
within the acoustic timescale, which is a few minutes in this event; afterwards, UV emission 
from the transition region could be governed by the pressure-gauge \citep{Qiu2013}, and 
therefore shields the slow-heating component that continuously heats the foot-point but at a low-rate. 
To discern these heating events, it is crucial to explore optically-thin 
UV observations with much higher spatial resolution, such as those provided by the recently
launched Interface Region Imaging Spectrograph \citep[IRIS; ][]{Depontieu2014}. Side by side with observational effort,
1d modeling is also necessary to comprehensively and coherently address both the corona and 
transition region physics in both the impulsive and gradual phases.

\subsubsection{What causes slow-heating?}
The modified heating profile suggests a scenario that energy release in a flare loop takes place 
in two phases, an intense impulsive heating phase and a gradual gentle heating phase. From the single pixel
statistics, the gradual heating may be present in the majority of flare loops. Flare loops are
formed by magnetic reconnection governed by Alfv\'enic timescales of order a few seconds, which is
consistent with the timescale of the impulsive heating. Then what mechanism produces the ubiquitous 
gradual gentle heating? A few plausible scenarios deserve further consideration. 
First, post-reconnection magnetic fields might not be relaxed to the lowest energy state within the Alfv\'en timescales, and post-flare
loops may still carry electric current with gradual Joule heating in the loops like
described by \citet{Parker1983, Schrijver2004}. This current dissipation scenario is equivalent to {\em slow reconnection}
operating with the classical resistivity, and the timescale of energy release is a few orders of magnitude 
longer than {\em fast reconnection}. If {\em fast reconnection} occurs on timescales of $\sim$10$^0$~s, the timescale
of the current dissipation may take place over a few tens of minutes to release comparable
amount of energy. Second, post-reconnection loops retract, or shrink \citep{Forbes1996}, 
under the magnetic tension force, and in this course releases energy. 
However, newly formed loops do not retract in vacuum and receive resistance from earlier formed low-lying loops,
generating slow-shocks \citep{Cargill1982, Cargill1983} with possibly elongated timescale of energy release.
In addition, the dynamic process during the flare may also trigger many magnetosonic waves \citep[see review by ][]{Aschwanden2006}. 
It is likely that the slow-heating process is governed by wave damping. The required slow-heating duration
of 20 - 30 min in the studied event is roughly comparable with damping timescales of different 
kinds of waves inside post-flare loops \citep[e.g. ][]{Wang2011}. 

\subsubsection{Limitation of the model}
The present study relies on a 0d model, which has enormous advantage to study a 
large number of loops or threads statistically \citep{Cargill2012b}, as has been done here. Questions can be raised 
regarding the fidelity of using the mean property approach to describe dynamically evolving flare plasmas. 
Admittedly this approach is not consistent with non-equilibirum physics during the 
impulsive heating phase \citep[such as the discussion in ][]{Qiu2013}, but the gradual phase of 
the long timescale appears to be governed by an approximate equilibrium,
validating the 0d approach. Plasma evolution in this gradual phase could be much less dependent on the details of the impulsive heating, as
demonstrated by numerous 1d models \citep[e.g. ][]{Winebarger2004} including the one employed here. 
There is certainly ample room for improvement. The datasets of this flare are also optimal to 
study plasma properties along the loops, which will constrain 1d models to help understand
the physics of heating. To calculate the synthetic loop emission, an ionization 
equillibrium is assumed. This is not necessarily true \citep[e.g. ][]{Shen2013}.
It has also been discussed that turbulence in flare loops may suppress thermal 
conduction \citep[][ and references therein]{Jiang2006, Wang2015}.
The impact of these effects on flare modeling can be examined in the future study.
Finally, we also note that any of the above-mentioned mechanisms, if 
responsible for the 
gradual heating, can be a natural consequence of dynamic plasma as well as
magnetic evolution of flare loops upon impulsive energy release, and therefore, 
may be addressed in a magnetohydrodynamic model beyond the 1d framework.

\acknowledgments We thank the referee for the careful review and insightful comments that help improve the paper.
We acknowledge the SDO mission for providing quality observations. This work is supported by the NASA grant NNX14AC06G.

\bibliography{local}

\clearpage

\begin{deluxetable}{ccccccccccc}
\tabletypesize{\scriptsize}
\tablecolumns{4} \tablewidth{0pt} \tablecaption{Duration and Cooling Times\tablenotemark{a}}
    \tablehead{
    \colhead{Bandpass}&
    \colhead{131} &
    \colhead{94} &
    \colhead{335} &
    \colhead{171}}

\startdata
obs. duration & 66 $\pm$ 53 & 74 $\pm$ 18 & 60 $\pm$ 22 & 12 $\pm$ 44 \\
syn. duration & 22 $\pm$ 12 & 33 $\pm$ 20 & 49 $\pm$ 28 & 10 $\pm$ 9 \\
              &           & 131 - 94    & 93 - 335    & 335 - 171 \\
obs. cooling  &           & 27 $\pm$ 26 & 58 $\pm$ 14 & 18 $\pm$ 19 \\
syn. cooling  &           & 13 $\pm$ 35 & 27 $\pm$ 36 & 32 $\pm$ 34 \\
\enddata
\tablenotetext{a}{Measured medium time in minutes; uncertainties are standard deviations.}
\end{deluxetable}

\clearpage

\end{document}